\pdfoutput=1
\documentclass{article}

\usepackage[preprint]{neurips_2019}

\usepackage[utf8]{inputenc} 
\usepackage[T1]{fontenc}    
\usepackage{hyperref}       
\usepackage{url}            
\usepackage{booktabs}       
\usepackage{amsfonts}       
\usepackage{nicefrac}       
\usepackage{microtype}      

\usepackage{xcolor}
\usepackage{graphicx}
\usepackage{amsmath,bm}
\usepackage{amssymb,mathtools}
\usepackage{algpseudocode}
\usepackage{algorithm}

\title{Emergence of Network Motifs \\
      in Deep Neural Networks}

\author{%
  Matteo Zambra\\
  Department of Civil, Environmental \\
  and Architectural Engineering\\
  University of Padova\\
  Padova, IT 35131 \\
  \texttt{matteo.zambra@studenti.unipd.it} \\
   \And
  Alberto Testolin\\
  Department of General Psychology and \\Department of Information Engineering\\
  University of Padova\\
  Padova, IT 35131 \\
  \texttt{alberto.testolin@unipd.it} \\
   \And
  Amos Maritan\\
  Department of Physics and Astronomy and\\
  Istituto Nazionale di Fisica Nucleare\\
  University of Padova\\
  Padova, IT 35131 \\
  \texttt{amos.maritan@unipd.it} \\
}

\begin{document}
\maketitle

\begin{abstract}Network science can offer fundamental insights into the structural and functional properties of complex systems. For example, it is widely known that neuronal circuits tend to organize into basic functional topological modules, called \emph{network motifs}. In this article we show that network science tools can be successfully applied also to the study of artificial neural networks operating according to self-organizing (learning) principles. In particular, we study the emergence of network motifs in multi-layer perceptrons, whose initial connectivity is defined as a stack of fully-connected, bipartite graphs. Our simulations show that the final network topology is primarily shaped by learning dynamics, but can be strongly biased by choosing appropriate weight initialization schemes. Overall, our results suggest that non-trivial initialization strategies can make learning more effective by promoting the development of useful network motifs, which are often surprisingly consistent with those observed in general transduction networks.
\end{abstract}

\section{Introduction}
The topological structure of complex networks can be characterized by a series of well-known features, such as the small-world and scale-free properties, the presence of cliques and cycles, modularity, and so on, which are instead missing in random networks \citep{newman2010, strogatz2001, caldarelli2010, newman2006, latora2017}. It has been shown that another distinguishing feature is the presence of so-called \emph{network motifs} \citep{milo2002}, which are recurring patterns of interconnections that might serve as building blocks for the evolution of more complex functional units \citep{lenski2003evolutionary, vespignani2003}.
In the words of \citep[Chapter~3]{alon2006}, one might thus hope to ``understand the dynamics of the entire network based on the dynamics of the individual building blocks''. In this respect, we can regard network motifs as basic structural modules which bear (in a topological sense) meaningful insights about the holistic behavior of the system as a whole.

Here we apply this perspective to the study of multi-layer (deep) neural networks, which are one of the most popular frameworks used in modern Artificial Intelligence applications \citep{lecun2015, goodfellow2016}. Despite the impressive performance achieved by deep networks in challenging cognitive tasks, such as image classification \citep{he2016deep}, automatic machine translation \citep{sutskever2014sequence} and discovery of sophisticated game strategies \citep{silver2016mastering}, such systems are still poorly understood \citep{montavon2018methods}. To quote \citep{saxe2019}, ``the theoretical principles governing how even simple artificial neural networks extract semantic knowledge from their ongoing stream of experience, embed this knowledge in their synaptic weights, and use these weights to perform inductive generalization, remains obscure''. The inscrutability of deep learning models mostly stems from the fact that their behavior is the result of the non-linear interaction between many elements, which motivates the use of network science techniques to reveal emergent topological properties \citep{testolin2018deep}.

The primary question we address in the present work is whether is it possible to observe the emergence of well-defined network motifs even if the initial (between layer) topology corresponds to a fully-connected graph, where each node (neuron) is connected to all nodes in the neighboring layers. The underlying assumption is that some traces of the learning dynamics will be nevertheless recorded in the final model topology in the form of basic functional modules, thus opening the possibility to relate local network properties with the functioning of the system as a whole. Furthermore, given that the objective of deep learning is to extract high-order statistical features from the data distribution \citep{testolin2016probabilistic}, we ask to what extent the final topology depends on intrinsic properties of the training data. To this aim, we systematically compare the motifs emerging in a deep network trained on two different synthetic environments, created according to different generative models.

\section{Methods}

\subsection{Neural network architecture}

A simple multi-layer feed-forward network was built and trained using the Keras deep learning framework\footnote{See \href{https://keras.io/}{https://keras.io/} for documentation.}. The external deep learning libraries and motifs mining software were used as provided, while the rest of the system was coded from scratch in Python 3.7.4\footnote{See \href{https://docs.python.org/release/3.7.4/}{https://docs.python.org/release/3.7.4/} for documentation.}. The complete source code is freely available at the GitHub repo \href{https://github.com/MatteoZambra/SM_ML__MScThesis}{https://github.com/MatteoZambra/SM\_ML\_\_MScThesis}.

The model consisted of four layers with a fully-connected architecture. The input layer contained 31 units, the two hidden layers were composed of 20 and 10 hidden units, respectively, and the output layers contained 4 units. Rectified linear units (ReLU) were used as activation function across all layers. Soft-max activation is instead used for the output units.

\subsection{Learning environments}

Inspired by the recent study of Saxe and colleagues \citep{saxe2019}, two synthetic sets of data were purposefully generated to embed different statistical structures, in order to investigate whether different environmental conditions would lead to the emergence of specific topological signatures. The first environment, encoded as a \emph{binary tree} data set, contained a hierarchical structure. The second environment, encoded as an \emph{independent clusters} data set, resolved in a more sparse and glassy statistical footprint, as shown in Figure~\ref{fig:tree_cov} and Figure~\ref{fig:clus_cov}. The structure of each data set will be briefly reviewed in the following; for a more detailed description the reader is referred to Appendix~\ref{app:data}. As customary in the machine learning literature, a data instance is thought of as a vector the entries of which are random variables, called features. It is in this case useful to think of these data generators as (probabilistic) graphical models, in which the relationships among the random variables involved are represented by means of edges between nodes, these latter encoding the just named random variables themselves. This yields an handy graphical formalism to handle this kind of models, especially the independent clusters case. 

In both the data sets, the number of classes in which the samples are divided is chosen to be four. 

\subsubsection{Binary tree data set} 

\begin{figure}
\centering
\includegraphics[width=0.62\linewidth]{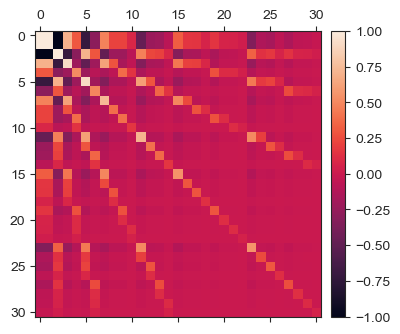}
\caption{Binary tree covariance matrix: the variables involved in the covariance computation are all the nodes of the tree structure, from the root node to the leaves.}
\label{fig:tree_cov}
\end{figure}

The binary tree data set is thought as a slight modification with respect to the generative model illustrated in \citep{saxe2019}. This data generator is designed to create data instances ensemble that display a hierarchical character, see Figure~\ref{fig:tree_cov}. Each generator run produces a data instance, features of which attain values among $\{-1, +1\}$. The initial feature (the root node) is sampled uniformly and its value diffuses through the tree branches. The underlying rationale is as follows: If the root node attains the value $+1$, then the left child inherits the $+1$ value and the right child (together with all its progeny) is assigned the value $-1$. Contrarily, if the root node happens to be $-1$, then the right child inherits the $+1$ value. From the root node children on, the criterion is probabilistic. One of the children of a $+1$ node inherits the same value with probability $\varepsilon$, that is a threshold that is set \emph{a priori}, in this case is set equal to $0.3$ (see Figure~\ref{fig:tree_explain}). Clearly, the smaller this threshold, less likely is the value to flip. In order to perform such a stochastic flip, for each node one samples a value $p$ uniformly distributed in $[0,1]$. For example: A node $i$ has the value $+1$. Given $\varepsilon$, sample $p \sim U(0,1)$. If $p > \varepsilon$, then the left child of $i$ (which has the node index equal to $2\times i + 1$, where the root node number is\footnote{And note that the node number is not the value stored in such node. This latter may be $-1$ or $+1$ while the former ranges from $0$ to the total number of nodes $N = 2^D -1$, $D = 1, \dots$ refers to the depth of the tree, that is how many nodes levels it has Note that linear array storage is used for the binary tree data structure.} $i = 0$) inherits the value $+1$ and the right child instead inherits $-1$. If otherwise $p \le \varepsilon$, the left child inherits $-1$ and the right child inherits $+1$. The features of each data sample are the collection of all the nodes in the tree structure for a given generator run, with the first feature being the root node, its right and left children the second and third features respectively and so forth.

The number of classes is set selecting a level in the tree: The root node (level $1$) identifies two axes of distinction, that is whether it value is $+1$ or $-1$. The data set fed to the network for our analyses is generated accounting for a level of distinction set to $2$.

\subsubsection{Independent clusters data set} 

\begin{figure}
\centering
\includegraphics[width=0.6\linewidth]{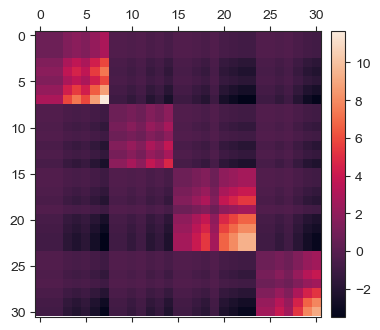}
\caption{Independent clusters covariance matrix.}
\label{fig:clus_cov}
\end{figure}

\begin{figure}
\centering
\includegraphics[width=0.5\textwidth]{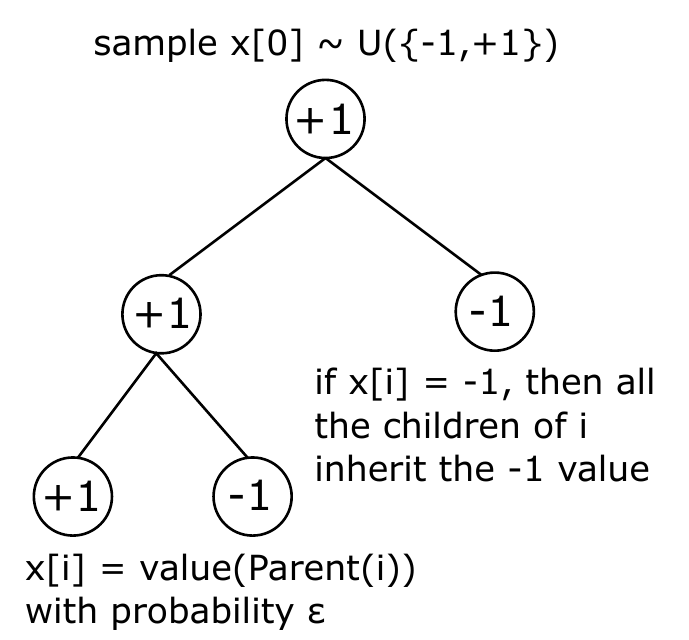}
\caption{Rationale behind the binary tree data sets creation.}
\label{fig:tree_explain}
\end{figure}

The independent clusters data set, on the other hand, is designed to endow the data instances ensemble with a block-diagonal statistical signature, as displayed in Figure~\ref{fig:clus_cov}. For such purpose, a simple graph is created as in Figure~\ref{fig:fcg}. For consistency with the number of features and classes with respect to the binary tree data set, the same number of random variables involved is chosen, and the number of independent groups is set according to the number of classes of the binary tree data set. In the full graph, as the embedding shown in Figure~\ref{fig:fcg}, some connections are gradually eliminated to reproduce the situation in Figure~\ref{fig:mg}. The rationale beneath is as follows: In a fashion similar to the \emph{simulated annealing} algorithm \citep{kirkpatrick1983}, a temperature schedule is set, where the initial temperature is chosen to be the reciprocal of the longest edge and similarly the final value is the reciprocal of the shortest edge

\begin{equation}
\left\{
\begin{array}{r c l}
T_0 &=& \left(\max_e\{e \in \mathcal{E} \}\right)^{-1} \\[2mm]
T_f &=& \left(\min_e\{e \in \mathcal{E} \}\right)^{-1}
\end{array}
\right.
\end{equation}

\begin{figure}
\centering
\includegraphics[width=0.5\textwidth]{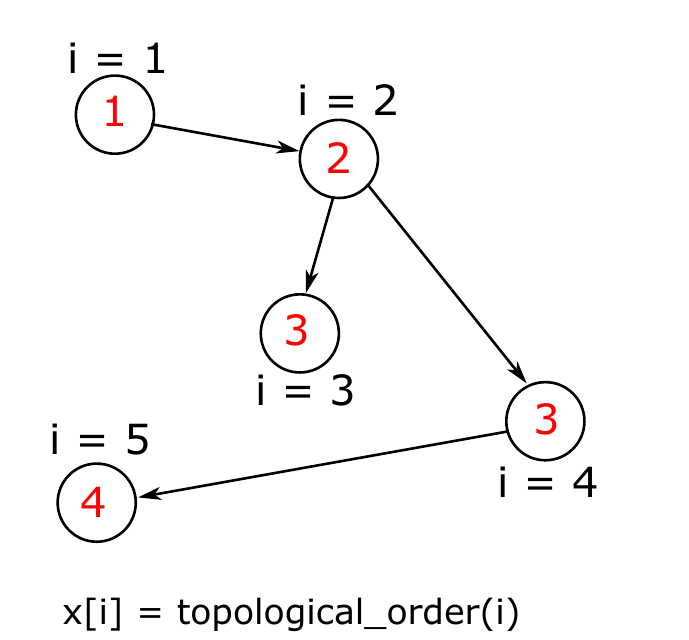}
\caption{Rationale behind the independent clusters data sets creation. The red labels represent the topological orderings of the respective nodes. The $i$ indices represent the nodes numbers.}
\label{fig:clus_explain}
\end{figure}

\begin{figure}
\centering
\includegraphics[width=0.7\textwidth]{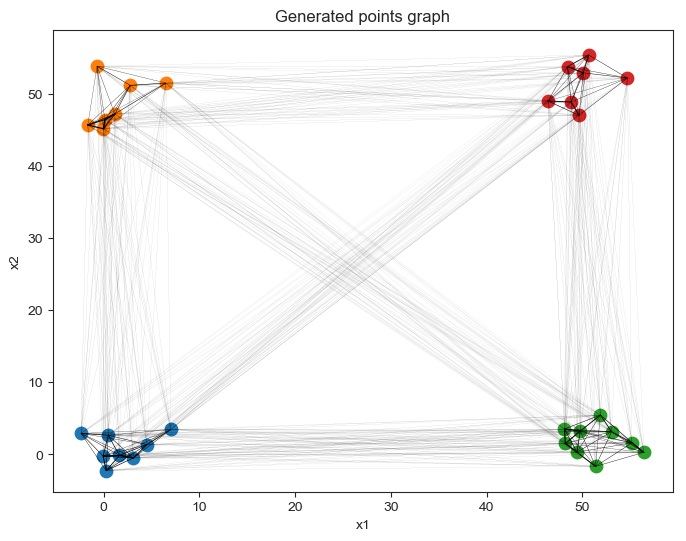}%
\caption{First stage (fully connected graph) of the data generating structure for the independent clusters case.}
\label{fig:fcg}
\end{figure}

\begin{figure}
\centering
\includegraphics[width=0.7\textwidth]{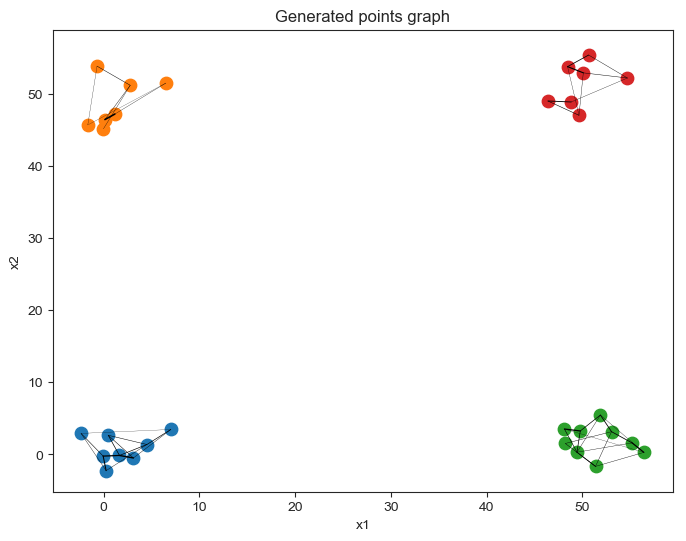}%
\caption{Second stage (\emph{Molten} graph) of the data generating structure for the independent clusters case. The connections between different groups are gradually eliminated in order to obtain independent graphs. Note that the geometric coordinates do not impact the values attained by the nodes. Coordinates are temporarily assigned in the creation stage for the purpose of visualization.}
\label{fig:mg}
\end{figure}

The schedule steps are equally spaced and the number of such steps is a parameters that has been fine-tuned in order to have the desired final scenario of disjoint, arbitrarily intra-connected groups. Note that by thus doing it could be that from a step to the next, shorter links may be erased while retaining longer ones, which were short enough to survive in the last step, hence melting schedule must be set sensibly. Unlike simulated annealing, temperature rises in this case and there is no thing such as an optimization purpose in this scheme. For each step, edges greater or equal than the inverse of the temperature value are deleted. Each of these four graphs is directed, in order to allow for a more efficient sampling. It is not strictly necessary to explicitly enumerate these groups. Each random variable (i.e. node value) is initialized to the value of $-1$. Subsequently, one selects randomly a group among the four available and the topological order of such nodes is set: The node with no incoming connection has topological order $1$. The nearest neighbors of this latter have topological order $2$, and so forth (see Figure~\ref{fig:clus_explain}). For the sake of simplicity, this data set is created in such a way that a random variable of the selected group bears as value its topological order. This translates in a straight-forward classification task, since all the data samples belonging to a class are equal: All the features of a data sample are set to $-1$, except those corresponding to the nodes of group selected, which are set equal to their topological ordering, in the scope of that specific group. See Appendix~\ref{app:data} for a thorough explanation. The construction of the structure depicted in Figure~\ref{fig:mg} is motivated by the necessity to dispose of a data set in which some random variables share a probabilistic relationship and some others are independent. Referring to Figure~\ref{fig:clus_cov}, the regions in the covariance matrix related to dependent variables are visible as the diagonal blocks, while the background appears more \emph{glassy}. This latter is not fully homogeneous but has rather a chessboard-like textured since the the numerical values attained by the non-chosen groups will somehow be related to the values of the chosen group. Each of such groups may display loops, but they do not necessarily have a tree structure. Recall that the connections are those retained after the ``melting'' process.

\subsection{Initial conditions}

Besides varying the statistical structure of the learning environment, we also investigated whether the emergence of different topological signatures could also be related to the use of different initialization schemes for the connection weights. To this aim, we adopted three different initialization schemes. In the first case, we used the classic ``Normal'' initialization method, where each connection weight is randomly sampled from a Gaussian distribution with zero mean and small variance, that is $w \sim \mathcal{N}(0.0, 0.1)$. In the second case we exploited the ``Orthogonal'' initialization method proposed in \citep{saxe2013}, where weight matrices of adjacent layers are constrained to be orthogonal. Finally, in the third case we exploited the popular ``Xavier'' (equivalently called ``Glorot'') initialization method \citep{glorot2010}, where the mean is zero and the variance of the Gaussian distribution is defined according to the number of connections of each layer, that is: 

\begin{equation}
\sigma^2 = \frac{k}{n_{\text{in}} + n_{\text{out}}}
\label{eq:glorot}
\end{equation}

$k$ depends on the activation non-linearity and $n_{\text{in, out}}$ are the number of incoming/outgoing connections, referring to one layer. Due to the modest number of units in this simple model, values $w$ are not exaggeratedly small in this case. The rationale behind these choices is the following: As explained hereafter, a pivotal role may be played by initialization schemes in the standpoint of topological features. Hence different of such strategies are tested. The normal initialization is the historically preferred one but in the course of developing of the deep learning discipline, some more clever schemes have been proposed. The ``Xavier'' scheme enjoy widespread popularity since it prevents back-propagated error gradients to vanish, a problem that afflicts deeper networks. Orthogonal matrices scheme has been proved to grant depth-independent training speed, which is desirable as the network becomes deep enough.

\subsection{Task and learning dynamics}

Stochastic (mini-batch) Gradient Descent was used to adjust the network weights. Learning rate was initially set to $0.01$, and then decayed using a factor of $10^{-6}$. Nesterov acceleration was added with momentum set to $0.6$.

The task accomplished by the network is multi-class classification. Referring to the learning environments described above, the binary tree domain is used as reference for the number of input features and output classes. The independent clusters domain, in turn, is generated in such a way to display the same number of features -- i.e. random variables -- and the same number of classes. The data sets are required to be homogeneous in terms of design matrix dimensions and items in the labels set, so that the network model is the same in both the cases: Once the parameters are initialized, the data structure containing these parameters and connectivity information is trained both on the binary tree and independent clusters environments.

The reason behind the presence of four output classes stems from the morphology of the binary tree data set. The choice of the level of detail set to $2$ means that one accounts for the classes identified by the equality of the data samples \emph{up to} the random variables of the second tree level \emph{plus} the respective outcomes; all the other variables may differ, thus having gathered in the same category data samples which display some difference in features. As explained in Appendix~\ref{app:data}, the choice of the level of detail in the tree hierarchical structure specifies the number of classes in which one can classify data samples. Then, for consistency, the independent clusters data set is purposefully designed to display four classes --i.e. four independent (not fully) connected groups of random variables. 

\begin{figure}
\centering
\includegraphics[width = 0.7\textwidth]{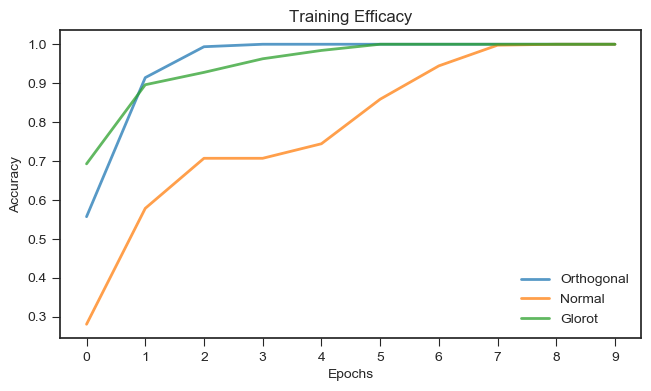}
\caption{Efficacy of initialisation schemes for the binary tree data set.}
\label{fig:eff_tree}
\end{figure}

\begin{figure}
\centering
\includegraphics[width = 0.7\textwidth]{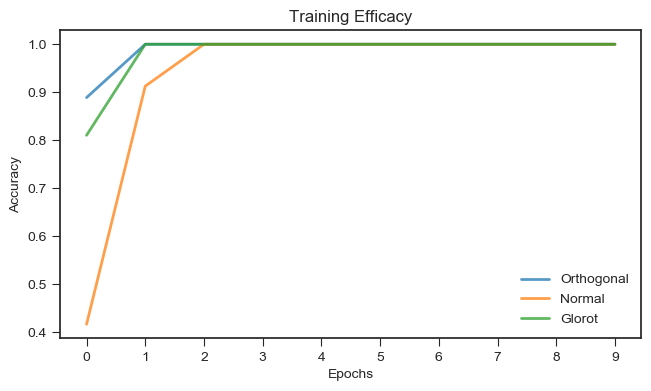}
\caption{Efficacy of initialisation schemes for the independent clusters data set.}
\label{fig:eff_clus}
\end{figure}

The tasks to be learned by the system are straight-forward, hence a simple model and algorithmic setup suffices to capture the problem complexity. The training accuracy peaks to top value of $1.0$ in few epochs, depending on the initialization scheme, as shown in Figures~\ref{fig:eff_tree} and \ref{fig:eff_clus}.

\subsection{Mining network motifs}

A concise overview of network motifs has been provided in the introductory paragraph. Here it is worth deepening the concept. Network motifs have been  defined as ``patterns of interconnection occurring in complex networks at numbers that are significantly higher that those in randomized networks'' \citep{milo2002}. To assess the statistical significance of such a pattern of connections, identifying a sub-graph, the $Z$-score (equivalently referred to as significance score) is defined, as the number of times a given motif appears in the real network with respect to the average number of occurrences of the same motif in an ensemble of random replicas of the original network (randomized networks are generated according to the local properties of the original network). It is basically a distance in units of standard deviations. In formulae it reads:

\begin{equation}
Z = \dfrac{N_{\text{real}} - <N_{\text{random}}>}{\sigma_{\text{random}}}
\end{equation}

\begin{figure}
\centering
\includegraphics[width=\textwidth]{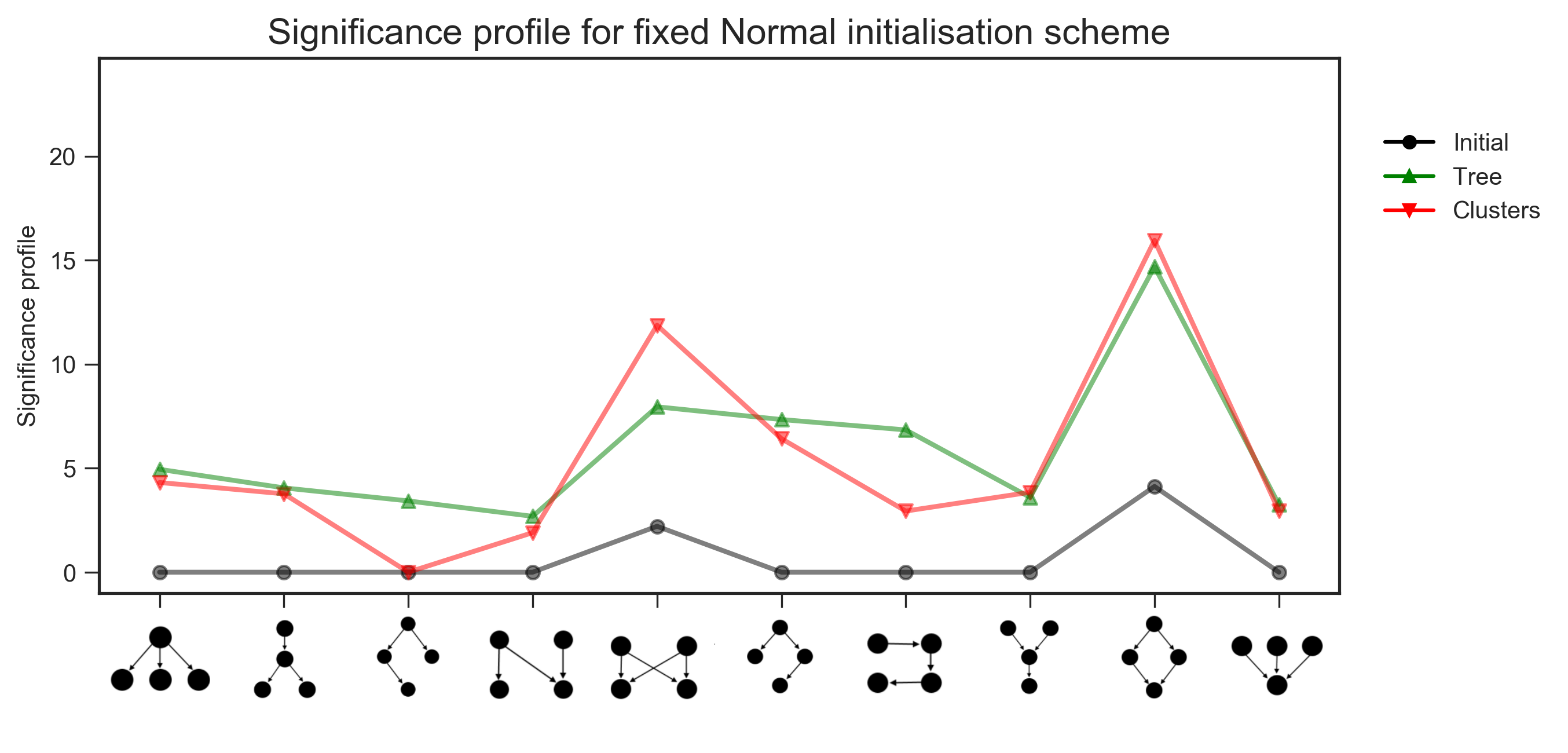}
\caption{Four-nodes motifs: significance profiles for the Normal initialization scheme.}
\label{fig:sp_init_normal}
\end{figure}

\begin{figure}
\centering
\includegraphics[width=\textwidth]{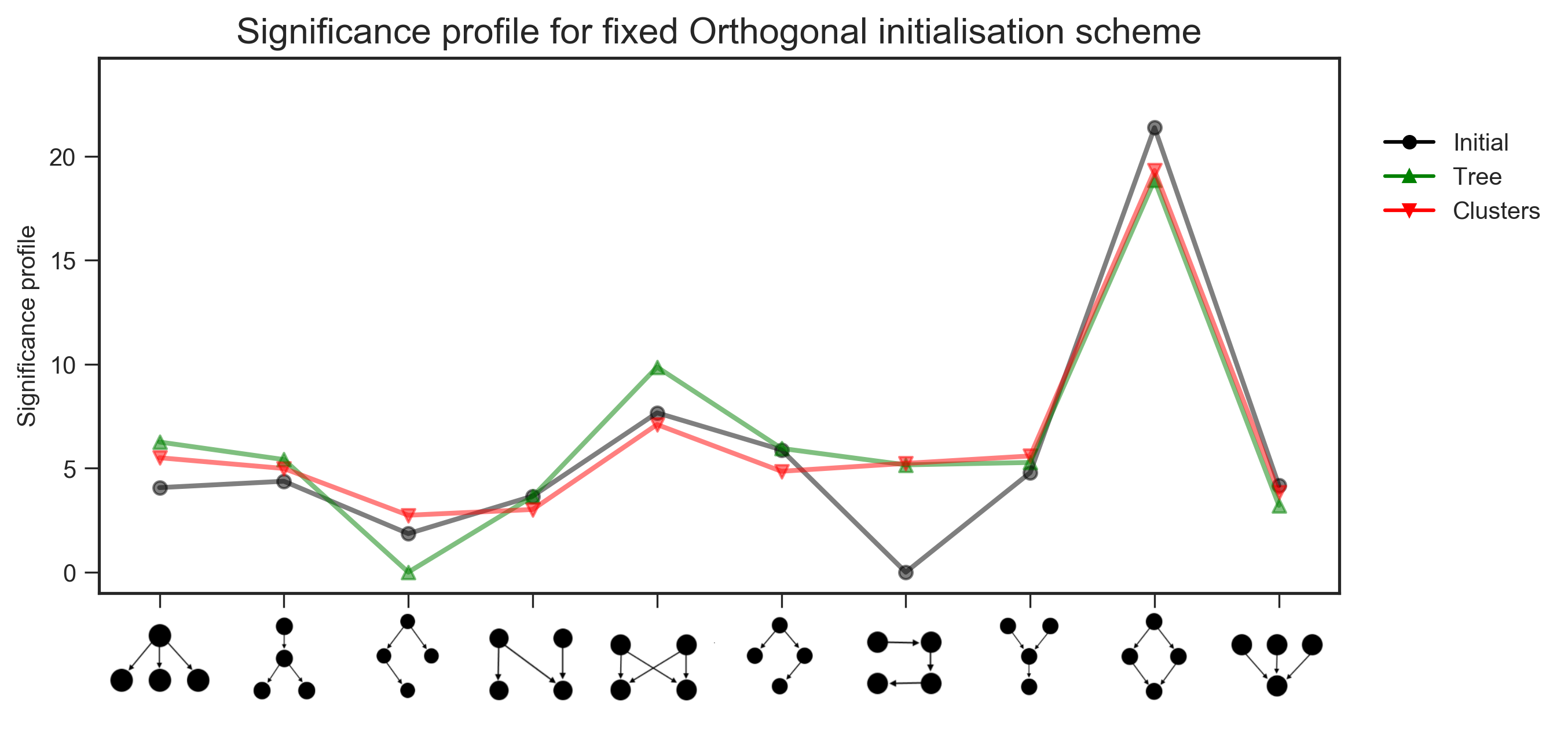}
\caption{Four-nodes motifs: significance profiles for the Orthogonal initialization scheme.}
\label{fig:sp_init_orth}
\end{figure}

\begin{figure}
\centering
\includegraphics[width=\textwidth]{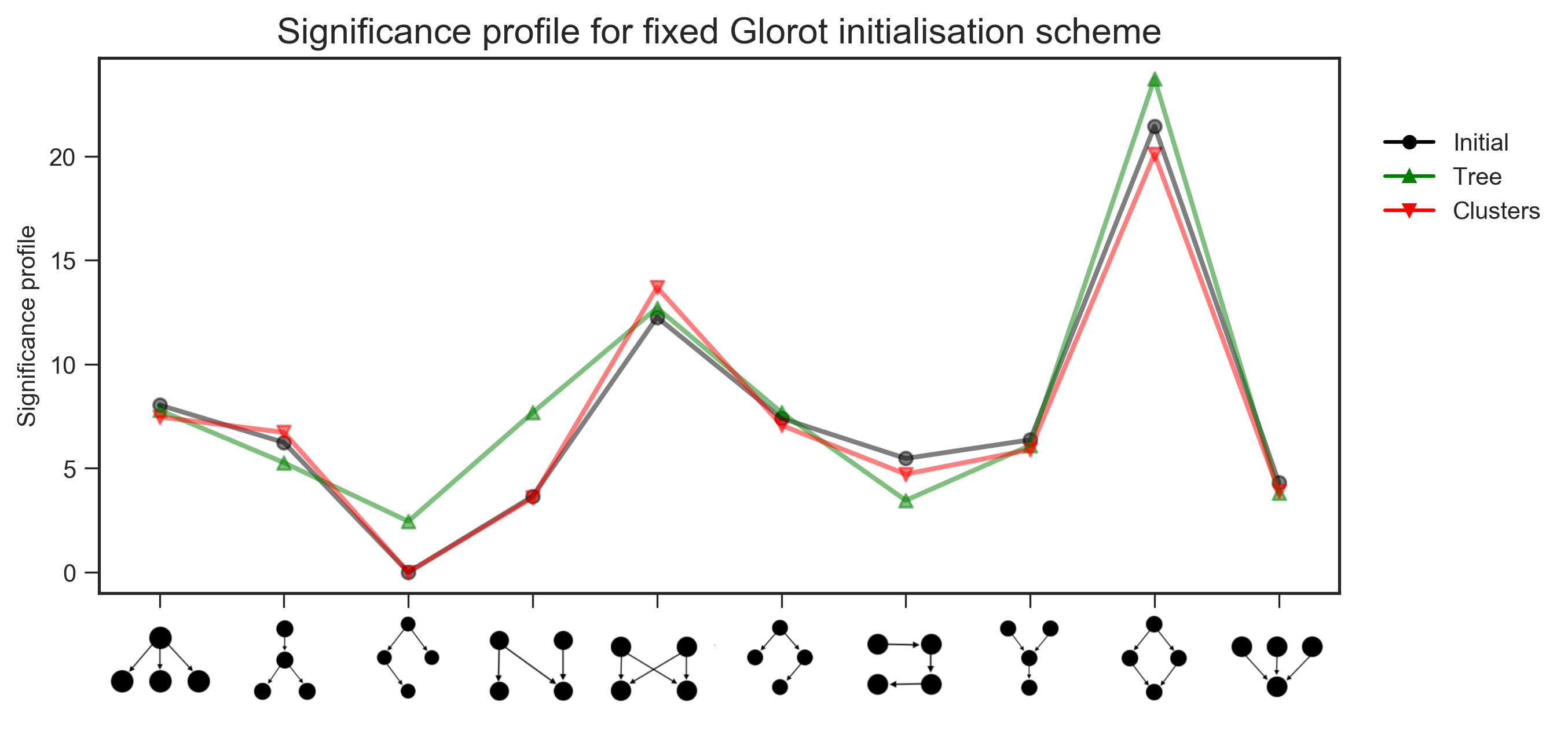}
\caption{Four-nodes motifs: significance profiles for the Glorot initialization scheme.}
\label{fig:sp_init_glorot}
\end{figure}

Each motif is then associated with its significance $Z$-score. It is visually explanatory to compare the $Z$-scores of all the motifs found. In Figures~\ref{fig:sp_init_normal}, \ref{fig:sp_init_orth} and \ref{fig:sp_init_glorot}, the $x$-axis gathers the motifs prototypes, the respective $y$ value is the significance score, obtaining a \emph{significance profile}. The reader should not be mislead by the trend-like aesthetic of such a graphical style, it does not represent any kind of temporally placed data. It is worth reporting here that it is customary to normalize the significance scores with respect to all the motifs found in a given network. In this way, significance profiles referred to different instances of complex network may be superposed \citep{milo2004}. However, in the present work normalization is avoided, since the network inspected (hence the size of the system) is the same for all the analyses. Non-normalized scores also allow to better understand the magnitudes of the detected effects.

Once training finished, model parameters were extracted and transposed in a proper graph data structure, so that network motifs mining can be carried on by means of an external software. The FANMOD\footnote{See \href{http://theinf1.informatik.uni-jena.de/motifs/}{http://theinf1.informatik.uni-jena.de/motifs/} for executable, sources, license and relevant papers.} motifs mining software was used to analyse the graph extracted from the model. The algorithmic standpoint covers a primal importance in the network motifs research thread. \citep{wernicke2006} outlines the FANMOD program, \citep{wernicke2006a} provides a detailed overview of its underlying algorithmic machinery. See also references therein and \citep{masoudi2012} for a comprehensive account on the historical benchmarks of the state-of-the-art advances.

The comparison between topologies before and after training is performed by means of the $Z$-scores of the motifs mined by this cited motifs mining tool.

\begin{figure}
\centering
\includegraphics[width=\textwidth]{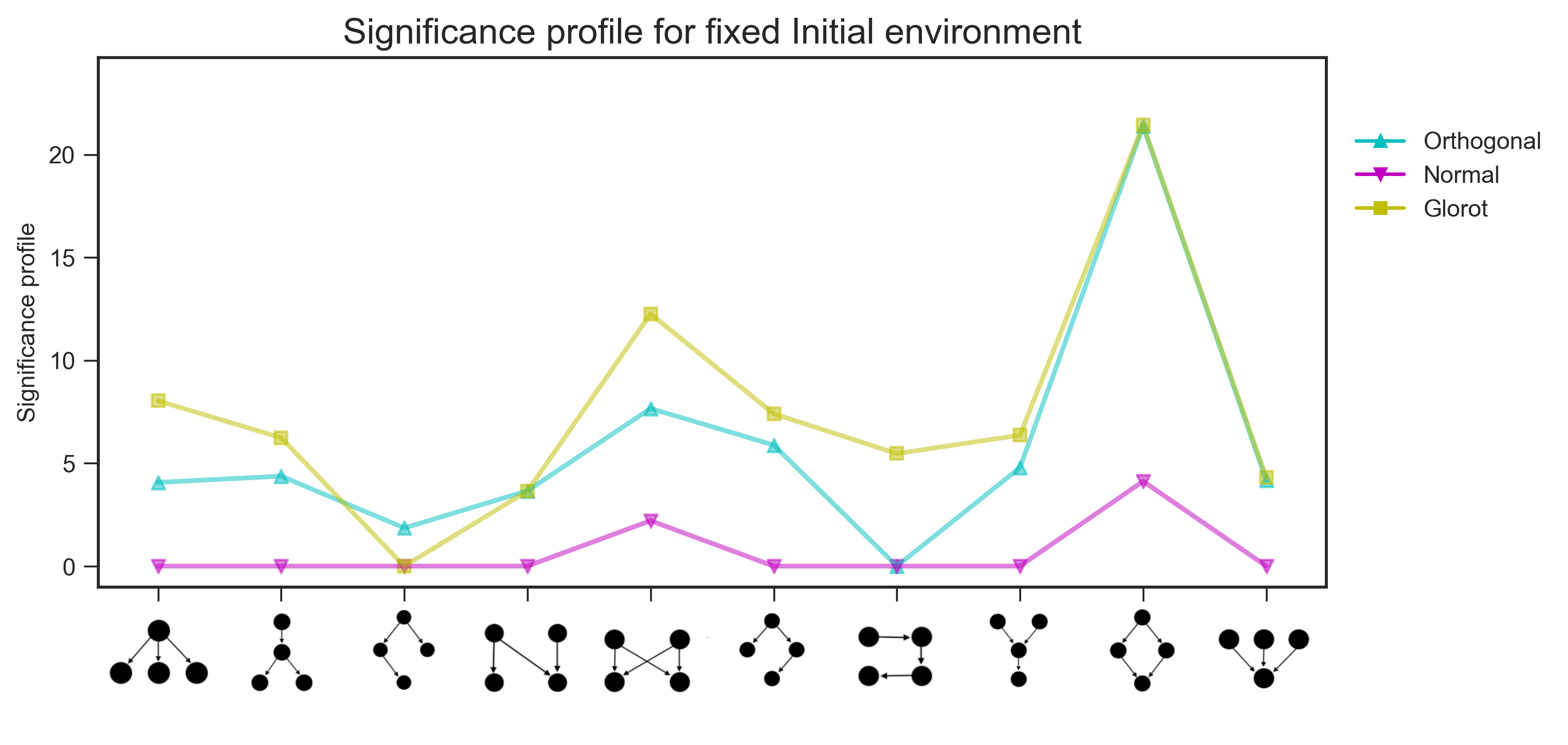}
\caption{Initial profile for all different initialization schemes.}
\label{fig:sp_env_init}
\end{figure}

\section{Results}

Now we define what we will mean with network motifs. In the present work, arrangements of four and six nodes are inspected. As we will point out, our analyses displayed an intriguing consistency with a class of biological networks, namely \emph{transduction networks}. Results for the four-nodes groups analyses are firstly presented. These allow to enjoy a broader perspective on the internal functioning of the system, compared to the biological counterpart. Six-nodes motifs results allow for a closer inspection of how learning environment and emergence of topological structures relate to each other. These results are less easily interpretable.

To motivate this choice of results presentation hierarchy, observe in Figures~\ref{fig:sp_init_normal}, \ref{fig:sp_init_orth}, \ref{fig:sp_init_glorot} that the significance profiles display a remarkable self-similarity. All the information one can extract from these plots is that there are two most significant motifs that are significantly present, regardless of the initialization scheme. Hence the inspection of four-nodes motifs yields this insight, but the details explaining how and why some topologies emerge are not available. Six-nodes motifs analyses are more informative regarding this aspect.

Since the relevant role of the initialization scheme appears to be pivotal, it is worth opening the results presentation with some remarks about the learning speed under different initial conditions.

\subsection{Learning efficacy}

As shown in Figures~\ref{fig:eff_tree} and \ref{fig:eff_clus}, the normal initialization scheme resolved in the slowest learning convergence. Instead, the orthogonal initialization scheme allowed convergence in few epochs. These findings suggest that initialization plays a crucial role in shaping learning dynamics: one possible explanation could be that the orthogonal and Xavier schemes impress a sharper fingerprint to the initial significance landscape of network motifs, as shown in Figure~\ref{fig:sp_env_init}. In other words, faster convergence toward the optimal set of connection weights might be promoted by biasing the initial set of network motifs. In fact, a sharper initial significance landscape is common in those initialization schemes which display a faster convergence.

\subsection{Biological analogy: neurons and protein kinases}

To deepen the contents of this section, the interested reader is referred to \citep[Chapter~6]{alon2006}, to which the topic and notation adopted in the following are inspired. The internal working of transduction networks is based on the cooperation between processing units, and the subsequent arrangement of those \citep{alon2007}. Sensing environmental stimuli, processing this information and eventually transcribing it to gene expression is done by passing this signal through a network whose units are \textit{protein kinases}. Such units play the role of nodes in the network, and interactions among those -- e.g. phosphorylation -- are the edges. Activity of these units are modelled by means of \emph{first-order kinetics}. The essential items are: 

\begin{itemize}
    \item the kinases of a first layer, the concentration of which is denoted as $X_j$, with $j = 1, \dots , n$;
    \item the target kinase of a second layer, the concentration of which is denoted $Y$;
    \item the rate of phosphorylation $r(Y) = Y_0 \, \sum_j v_j X_j$, being $v_j$ the rate of kinase $X_j$.
\end{itemize}

Call $Y_0$ and $Y_p$ the concentration of un-phosphorylated and phosphorylated kinase $Y$ respectively. The concentration of kinase $Y$ remains constant, that is $Y_0 + Y_p = Y$. Then the rate of change of activated kinase $Y$ is given by the difference between the rate of phosphorylation $r$ and the rate of de-phosphorylation of the same kinase $Y$, at a rate $\alpha$. In formulae

\begin{equation}
\left\{
\begin{array}{r c l}
r(Y) &=& Y_0 \, \sum_j v_j X_j \\[1mm]
Y &=& Y_0 + Y_p \\[1mm]
\dfrac{dY_p}{dt} &=& r(Y) - \alpha Y_p
\end{array}
\right.
\end{equation}

\begin{figure}[htb]
	\centering
	\includegraphics[width=0.5\textwidth]{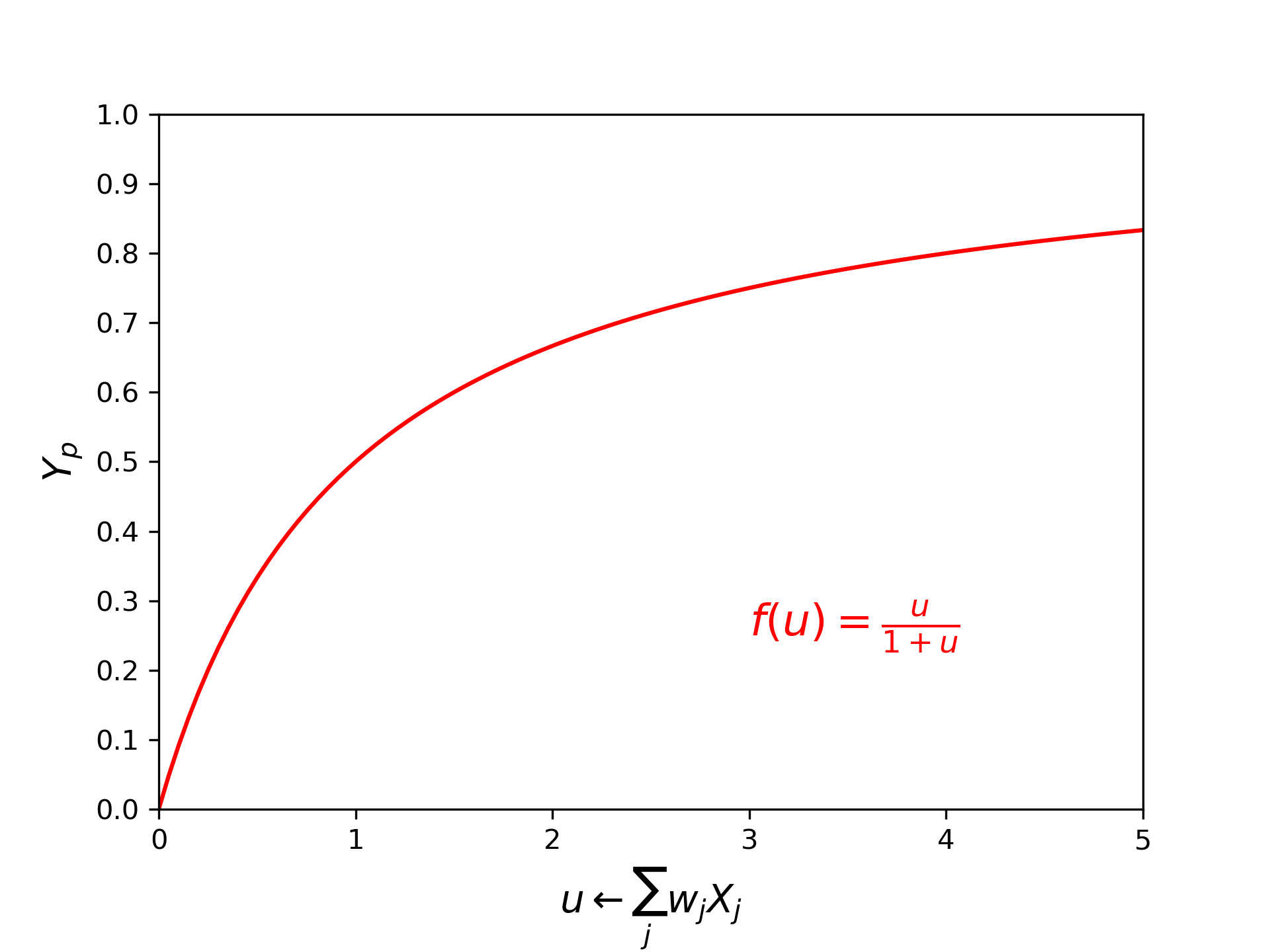}
	\caption{Behavior of the threshold function which quantifies the activity of target kinase, that is $Y_p$, as a function of the weighted sum of the input signals. A sensible value of the input weighted sum for the target unit to show activity is assumed to be approximately $1$ \citep[Page~109]{alon2006}. Would one not to make such an assumption, then the expression of the hyper-locus referred to in the main text is more generally $\sum_j w_j X_j = k$.}
	\label{fig:thre}
\end{figure}

Referring to the case of steady state dynamics, straight-forward calculations yield that the concentration of active $Y$ is non-linearly proportional to the weighted sum of the inputs $X_j$, as depicted in Figure~\ref{fig:thre} 

\begin{equation}
Y_p = \dfrac{\sum_j w_j X_j}{1 + \sum_j w_j X_j}
\end{equation}

where $w_j = v_j \alpha^{-1}$. A sensible threshold value for this weighted sum is thought to be $1$ approximately. After the value of the input exceeds $1$, the target kinase activity start to be sensible. Now assume that this simple model involves $m$ target kinases. Then

\begin{equation}\label{W}
\sum_{j = 1}^{n} \, w_{ij} X_j = 1 \quad i = 1 , \dots , m
\end{equation}

identifies the hyper-plane in the space of the inputs that excludes regions of high and low activity, depending of the connection strengths. 

It happens that by stacking more of such three-nodes modules ($n$ input signals from the kinases $X_1, \dots, X_n$ and the target unit $Y$), one can obtain complex geometries of the activity region in the input space. It is shown that the motifs encountered most often in transduction networks are the so-called \emph{diamonds} (the ninth from the left on the horizontal axis in Figure~\ref{fig:sp_init_normal} and following) and \emph{bi-parallel} (the fifth in the same way). These motifs match those found in out analyses.

The analogy with binary classification is hinged on the creation of the hyper-plane. Assume that the weighted input is $u = \sum_j w_{ij}X_j$, the numerical value of which is known. In order to determine whether the unit $Y$ is active, one needs to compare the input $u$ with the hyper-locus that identifies the regions of activity. Assume that the target unit activates once the threshold $1$ is exceeded, then

\begin{itemize}
    \item If $u \ge 1$ then target unit $Y$ activates and propagated the signal forward in the system to a third layer. But
    \item If $u < 1$ then $Y$ is not sufficiently triggered to propagate the signal, that is, to phosphorylate the next unit.
\end{itemize}

The hyper-space $W \bm{X} \ge \bm{1}$, $\bm{1} = \{1\}^n$, identifies the set of weights and activities such that the target units is activated. The subtlety in this analogy is that the weights $W$ cover a relevant role too: In transduction networks change of such weights is subjected to regulatory mechanisms or evolutionary pressure \citep{alon2006}, and in the process of gene expression transcription these weights values are given. They do not play the role of adjustable parameters in such a way to minimize a given error metric. In neural networks, on the other hand, weights adjustment is pivotal in the learning process, and such variations are performed on a faster timescale.

\begin{figure}
\centering
\includegraphics[width=0.7\linewidth]{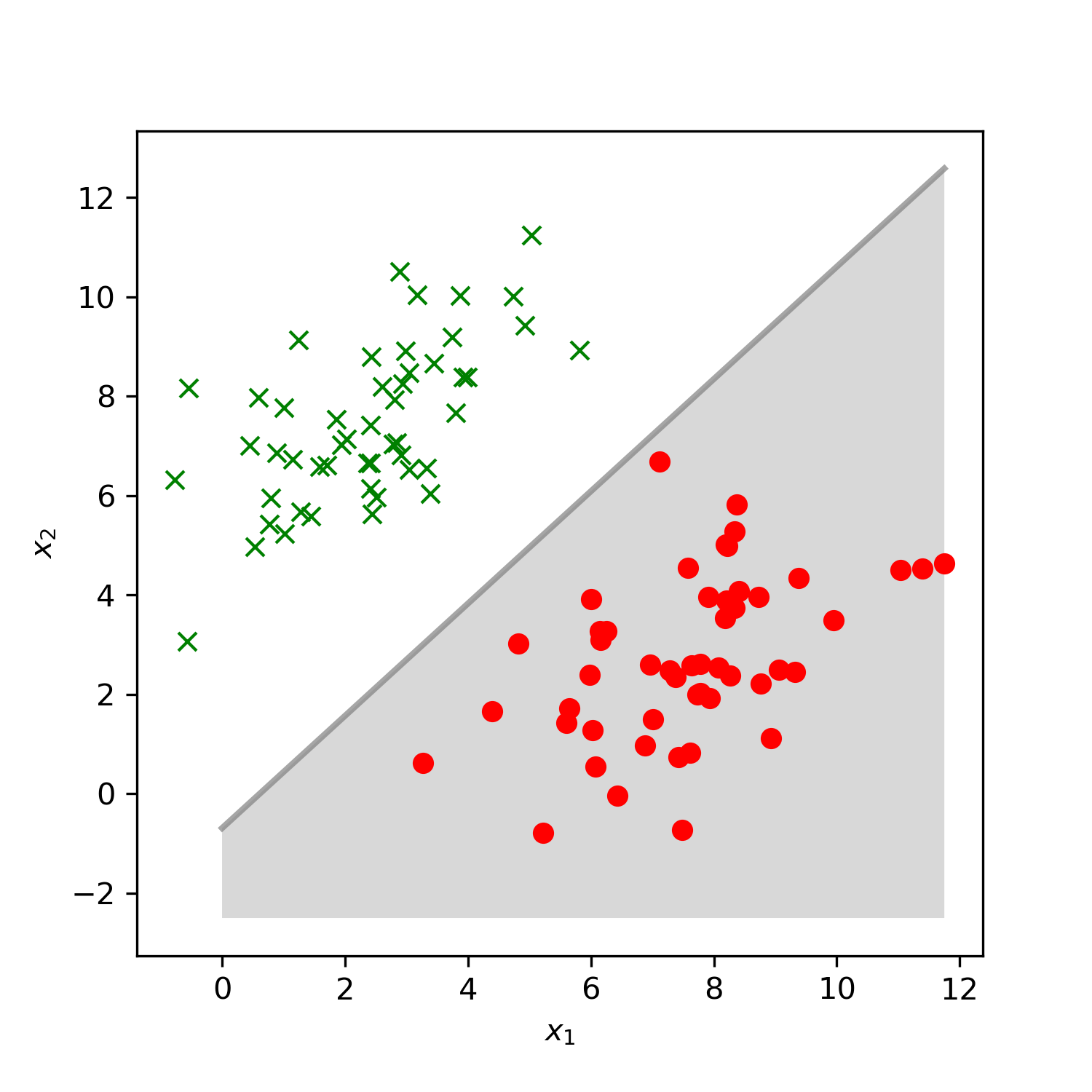}
\caption{Exclusion hyper-locus of a single neuron. The $x_1$ and $x_2$ coordinates represent the features of a fictitious data vector, featuring two random variables, in a case of linear separability. Here two input neurons map the input features to a binary label.}
\label{fig:lsc}
\end{figure}

In neural networks one encounters a similar scenario: A neuron is fed with an array of incoming signals, coming from the activities of the previous layers neurons. The weighted sum of these signals is added to an activation threshold value, called bias. The resulting value undergoes a non-linear transformation. In this way it is possible to identify an hyper-plane in the input space that separates the input patterns of signals, as in Figures~\ref{fig:lsc} and following ones. The ``state equations'' of a simple one-hidden-layer network are the following:

\begin{equation}
\left\{
\begin{array}{r c l}
\bm{h} &=& f \left( \bm{x} W^{(1)} + \bm{b}^{(1)} \right) \\[2mm]
\bm{y} &=& f \left( \bm{h} W^{(2)} + \bm{b}^{(2)} \right) \\
\end{array}
\right.
\end{equation}

with

\begin{equation*}
\begin{array}{l}
\bm{x} \in \mathbb{R}^{N_{\text{input}}}, \quad \bm{h} \in \mathbb{R}^{N_{\text{hidden}}}, \quad \bm{y} \in \mathbb{R}^{N_{\text{output}}} \\[2mm]
\bm{b}^{(1)} \in \mathbb{R}^{N_{\text{hidden}}}, \quad \bm{b}^{(2)} \in \mathbb{R}^{N_{\text{output}}} \\[2mm]
W^{(1)} \in \mathbb{R}^{N_{\text{input}} \times N_{\text{hidden}}}, \quad W^{(2)} \in \mathbb{R}^{N_{\text{hidden}} \times N_{\text{output}}}
\end{array}
\end{equation*}

Suppose that this network has one hidden layer with $N_{\text{hidden}}$ units, $N_{\text{input}}$ input units, $N_{\text{output}}$ output units and $f(\cdot)$ is a generic non-linearity. In the framework of neural networks these functions are generally monotonically increasing, as for example the ``logistic sigmoid'' $\sigma(x) = (1 + \exp(-x))^{-1}$. The vectors $\bm{b}^{(k)}$ represent the activation thresholds of both the hidden units and output units -- also called \emph{biases} and the matrices $W^{(k)}$ are the connection weights, $k = 1,2$. Note that the above mentioned state vectors are intended as row vectors.

Here the hyper-plane is identified by the weighted sum in the arguments of $f(\cdot)$, which purpose is to capture higher-order correlations in the input features and the composition of many non-linear blocks allows to synthesize high-level abstraction of the domain \citep{lecun2015}. Figures~\ref{fig:lsc} and following ones give a visual idea of the hyper-planes composition and the result in terms of decision boundary geometry. Signal flow in the system, from the input layer units to the output nodes, is strictly feed-forward and once the guessed label is observed in this latter layer, it is compared with the ground truth. Based on the mismatch, model parameters (connection weights and node biases) are adjusted, in such a way to minimize the prediction error, back-propagating such error in a reverse way along the layers constituting the network \citep{rumelhart1986}.

\begin{figure}
\centering
\includegraphics[width=0.7\linewidth]{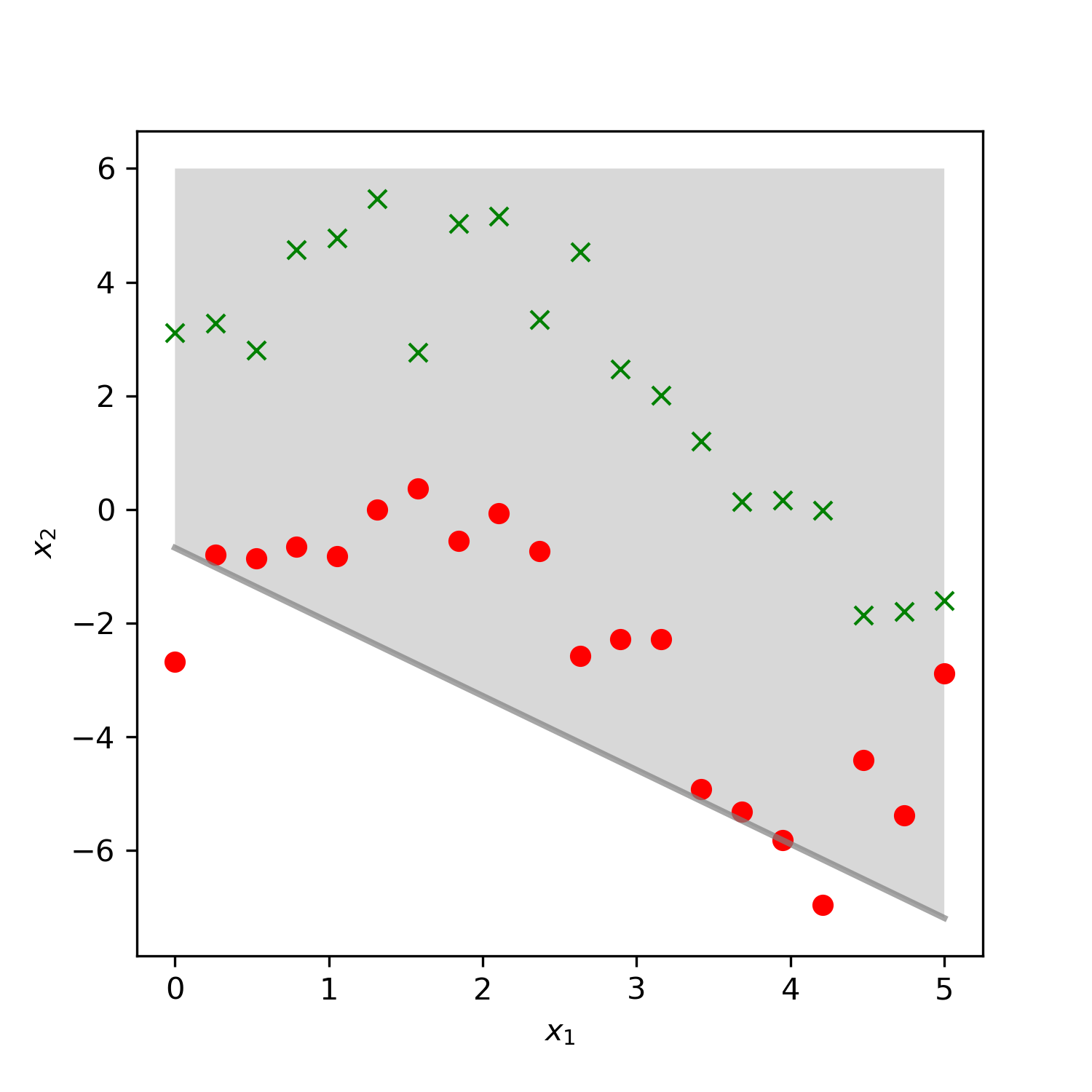}
\caption{Exclusion locus of the first triad.}
\label{fig:tc1}
\end{figure}
	
\begin{figure}
\centering
\includegraphics[width=0.7\linewidth]{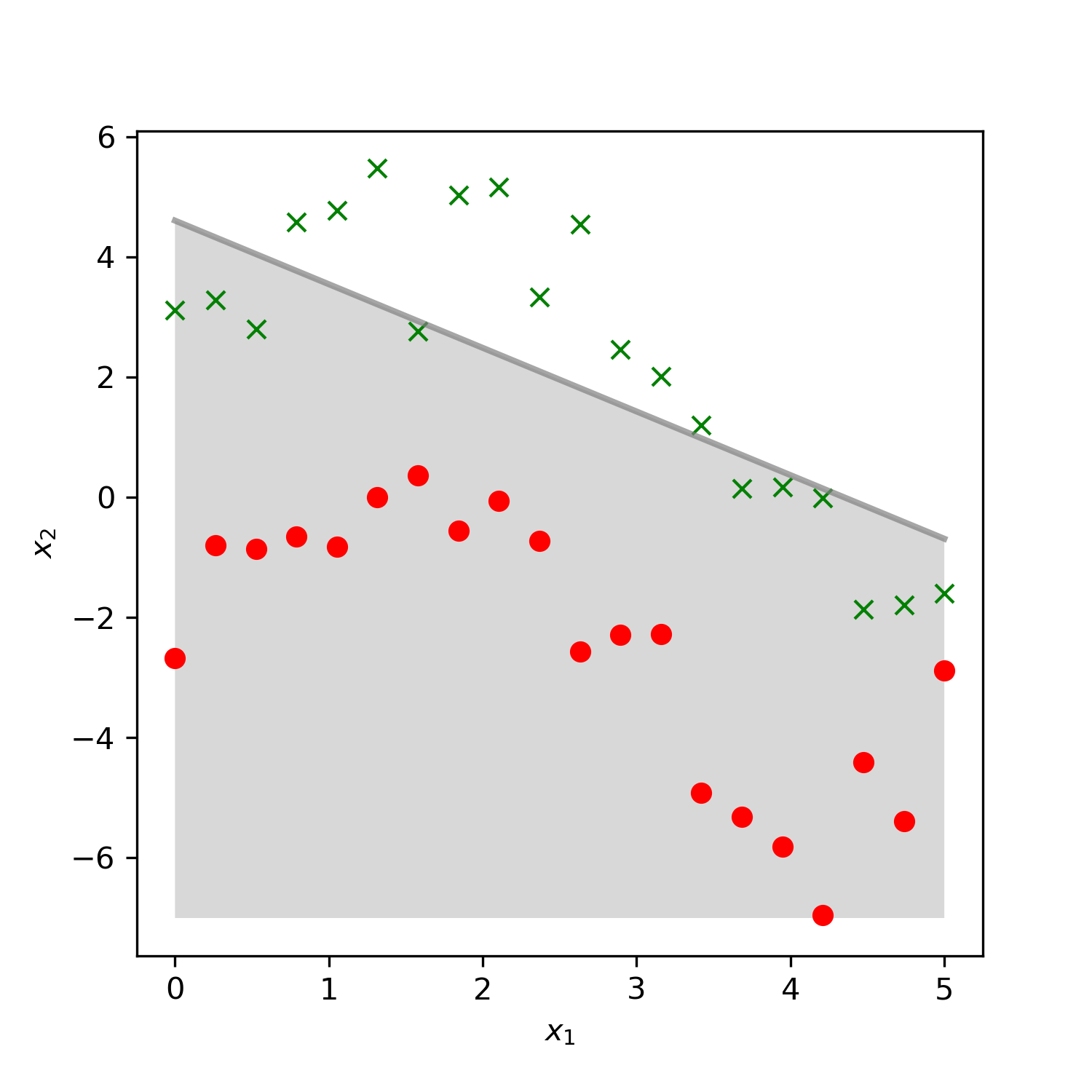}
\caption{Exclusion locus of the second triad.}
\label{fig:tc2}
\end{figure}
	
\begin{figure}
\centering
\includegraphics[width=0.7\linewidth]{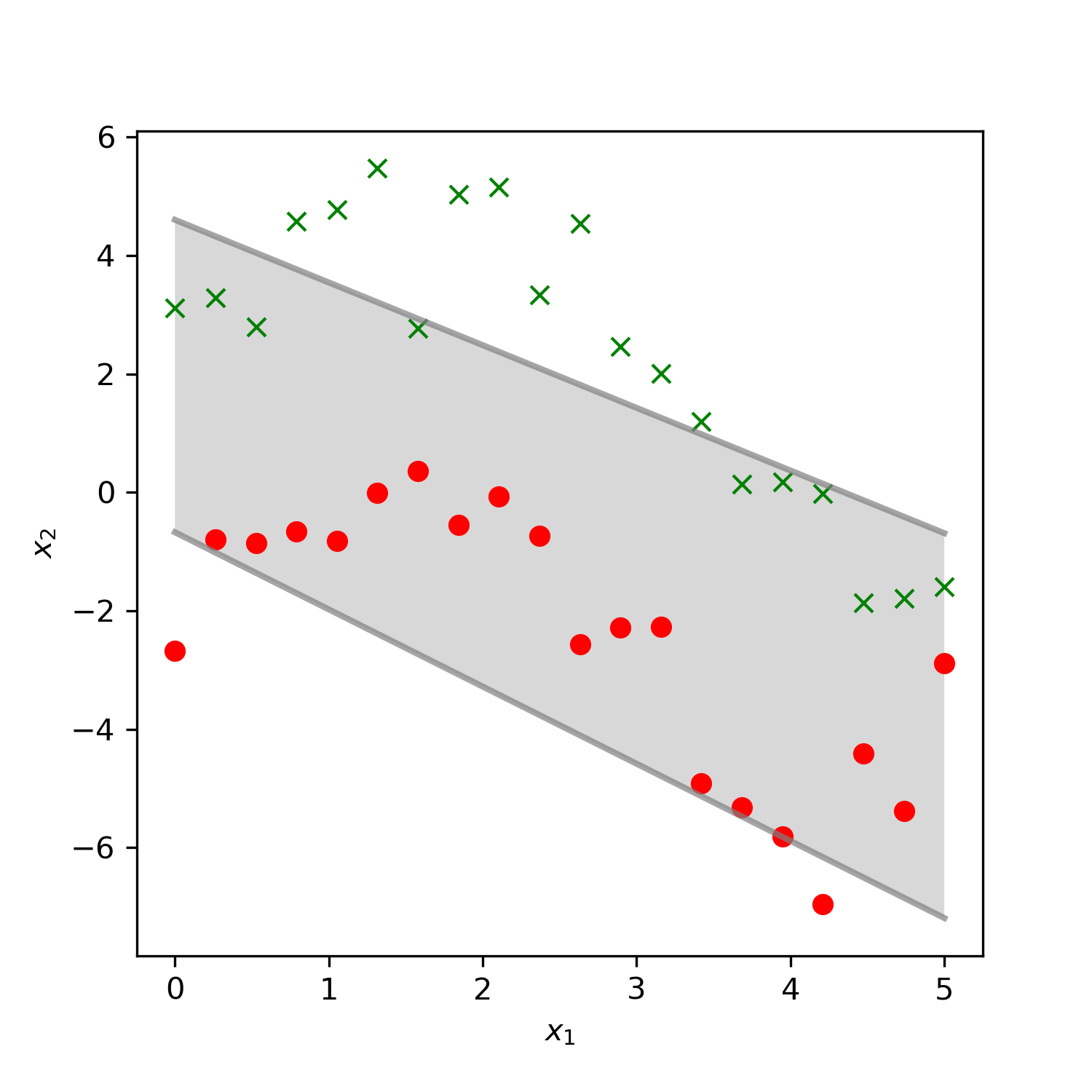}
\caption{Combination of the two hyper-loci, where the joint contribution of two  loci can allow one to go beyond the case of binary classification and linear separable classes, once the problem becomes more complex. Of course \emph{hyper-} is used with no loss of generality. In this example the decision boundary can be safely called \emph{plane} or line.}
\label{fig:stack}
\end{figure}

Given these analogies between biological transduction networks and artificial neural networks, it is thus legitimate to hypothesize that information processing in both classes of systems might be carried out using similar computational structures. \citep[Chapter~6]{alon2006} argues that ``Multilayer perceptrons allow even relatively simple units to perform detailed computations in response to multiple inputs. The deeper one goes into the layers of perceptrons, the more intricate the computations can become''. If one thinks of \emph{intricate computations} as the computation of appropriate decision boundaries, then this task is precisely what is accomplished by multi-layer perceptrons. Individual neurons (absorbing an arbitrary length input) could only discriminate two classes, as in Figure~\ref{fig:lsc} (in this case one has only two input features), but stacking together multiple layers of neurons allows to create more intricate and complex decision loci in the input space, as in Figure~\ref{fig:stack}. Panels~\ref{fig:tc1}, \ref{fig:tc2} and \ref{fig:stack} refer to the combination of two triads as in \ref{fig:lsc}, assembled so to form a simple neural network with two input neurons and two output units, with no hidden layers. \ref{fig:tc1} is the exclusion locus of the triad formed by the input units and the first output unit, \ref{fig:tc2} analogously refers to the triad in which the output unit involved is the second. This trivial example shows how hyper-planes designed by simple groups of units arrange to identify less obvious exclusion hyper-subspaces; and note that stacking more layers one can go beyond straight lines.

In our analyses of four-nodes motifs, we discovered a set of basic structures that might indeed support information processing, in the way explained above. Interestingly, several of such motifs are consistent with those commonly found in biological transduction networks, suggesting a potential overlap of computational mechanisms. The fundamental feature of the analogy is the identification of an hyper-plane which \emph{classifies}, \emph{distinguishes} the nature of a given input -- which comes as a weighted sum, in both transduction and neural networks. In the former case, it is documented that such a dynamics relies on the presence of network motifs. Our analyses show that the most significant motifs detected in neural networks match those commonly found in transduction networks, if four-nodes motifs are accounted for.

\begin{figure}
\centering
\includegraphics[width=0.8\textwidth]{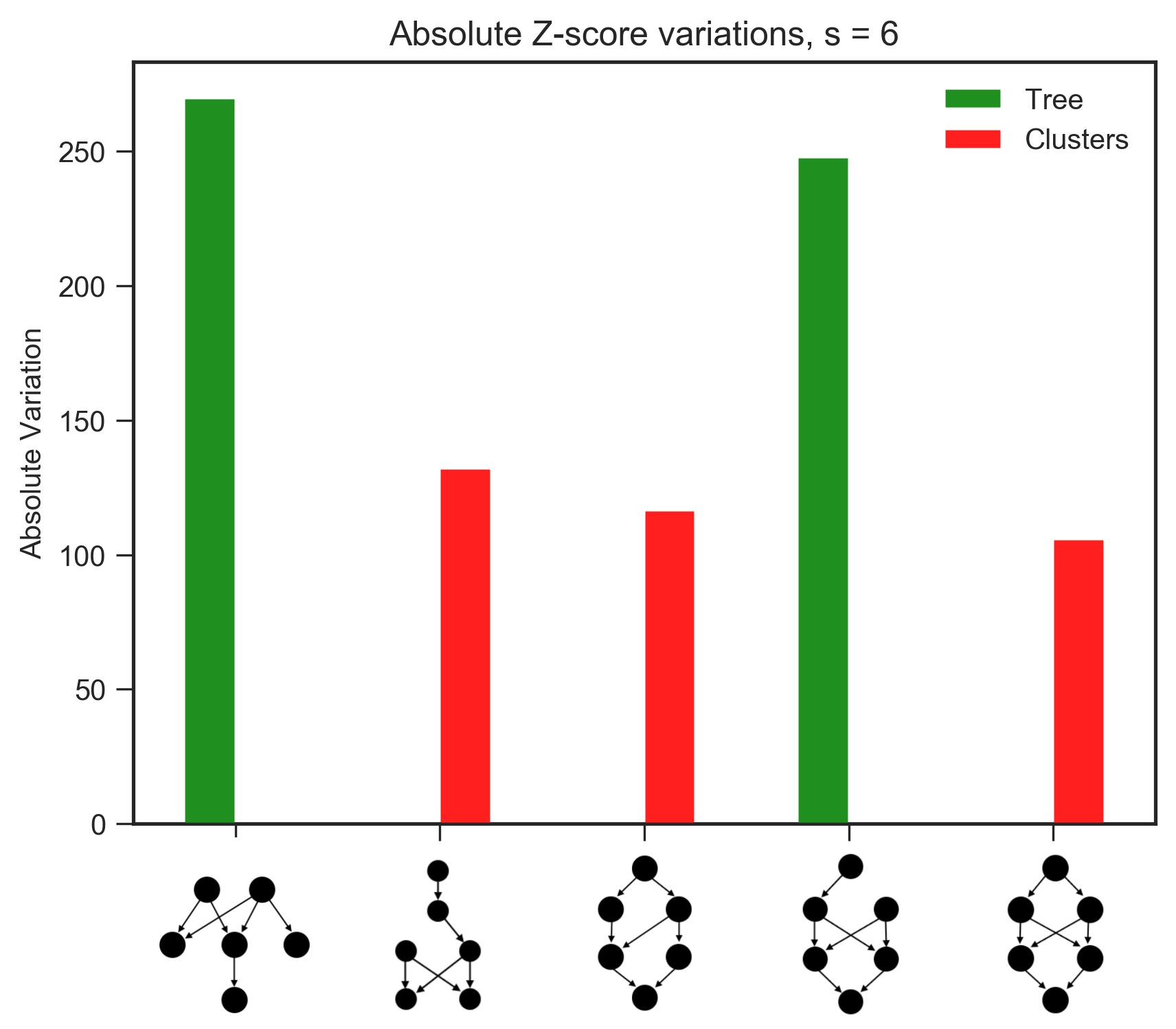}
\caption{Six-nodes motifs for the Normal initialization scheme. Total $Z$-score variations accounting for the difference in significance before and after training. Figure refers to most significant motifs, having analysed the weighted graph from the model.}
\label{fig:zsc_variation_normal_6nm}
\end{figure}

\begin{figure}
\centering
\includegraphics[width=0.8\textwidth]{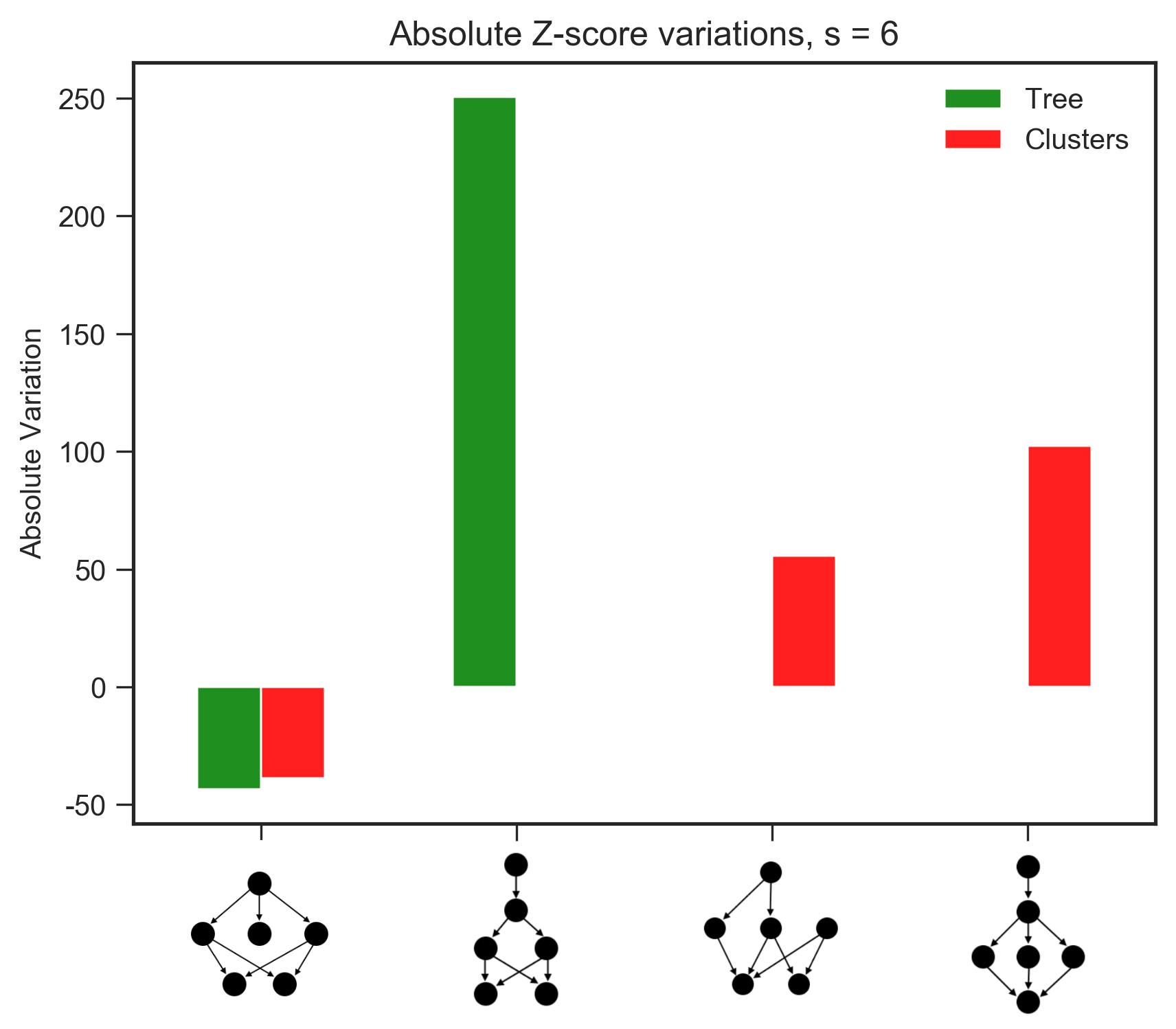}
\caption{Six-nodes motifs for the Orthogonal initialization scheme. Total $Z$-score variations accounting for the difference in significance before and after training. Figure refers to most significant motifs, having analysed the weighted graph from the model.}
\label{fig:zsc_variation_orth_6nm}
\end{figure}

\begin{figure}
\centering
\includegraphics[width=0.8\textwidth]{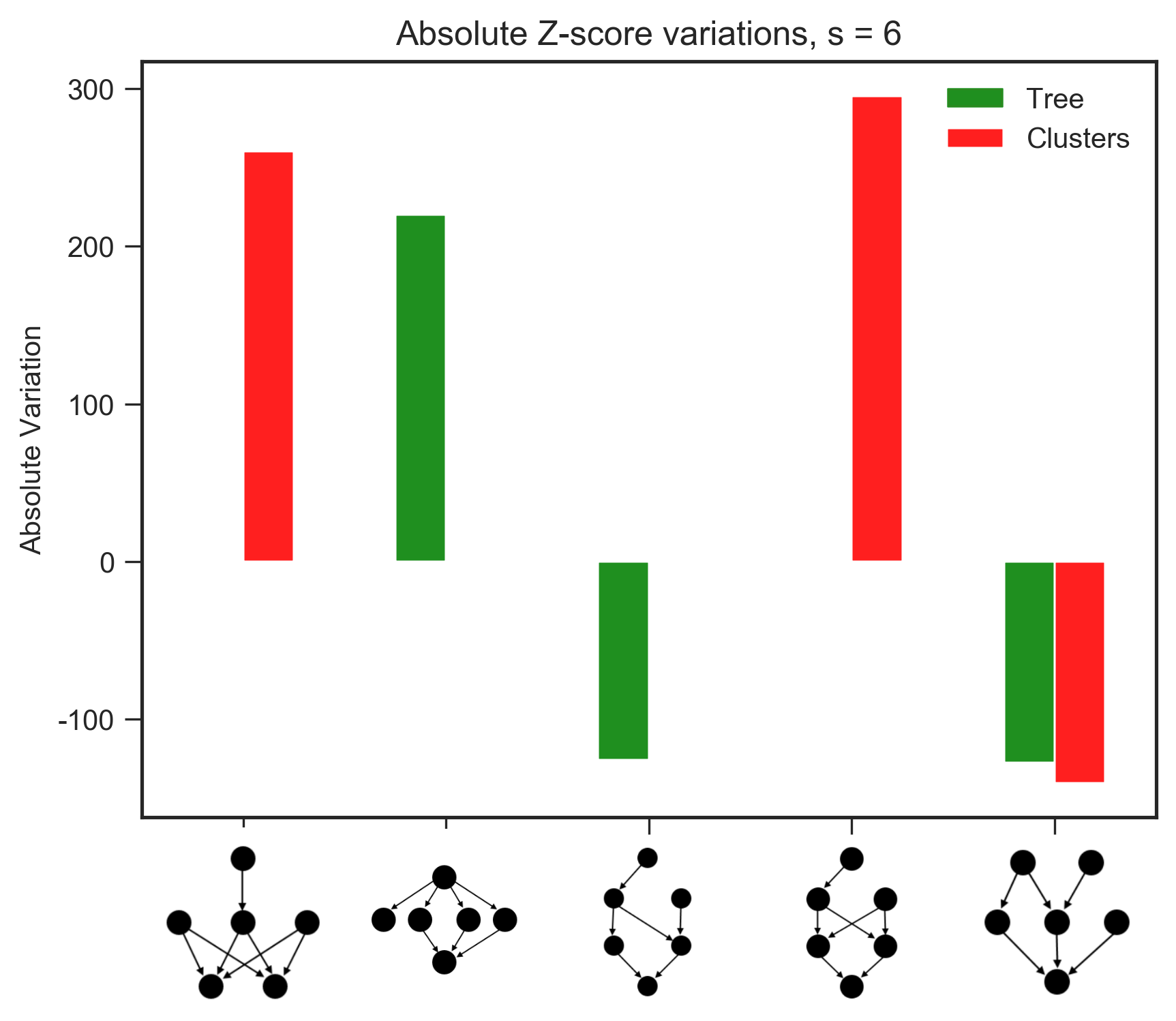}
\caption{Six-nodes motifs for the Glorot initialization scheme. Total $Z$-score variations accounting for the difference in significance before and after training. Figure refers to most significant motifs, having analysed the weighted graph from the model.}
\label{fig:zsc_variation_glorot_6nm}
\end{figure}

\subsection{Sensitivity to learning environments and initial conditions}

The second highlight itself encapsulates two distinct aspects: On the one hand it is possible to appreciate the exclusiveness of the characterizing motifs found for each arrangement of learning environments, thus it may seem that the domain influences the emergence of particular topologies. On the other hand, results are less obvious. Referring again to Figures~\ref{fig:sp_init_normal}, \ref{fig:sp_init_orth} and \ref{fig:sp_init_glorot} and~\ref{fig:sp_env_init}, it is not clear the extent in which the emergence of the most significant motifs is \emph{spontaneous} or biased by the initial profile (black lines). Orthogonal matrices and Xavier schemes, by their design, sample larger parameters values, hence the model configuration is conditioned by the initial, albeit random, weights landscape.

Our current results do not provide a definitive explanation: while it may seem that different motifs emerge in response to different initial and learning environments, is not clear why and when a topology is observed. The \emph{multilayer-perceptron} motif and its variations appear consistent through the different scenarios. In \citep[Chapter~6]{alon2006} it is argued that such a structure\footnote{In the cited work, it is the five-nodes structure that is inspected. The same as shown here but without the topmost node. It may change little however.} can be viewed as a combination of \emph{diamond} four-nodes motifs. 

The instances of that multilayer-perceptron motif appear rather blurred: One or more edges are not present in the other instances, see for example the third and fourth motifs in panels~\ref{fig:zsc_variation_normal_6nm} and \ref{fig:zsc_variation_glorot_6nm}. This is thought to be a side effect of the noise injected by the training algorithm. In principle, learning dynamics equations may seem to be deterministic, but due to the choice of the Stochastic Gradient Descent algorithm, it is inevitable for the end results to be subjected to noise and randomness. Thus, albeit one may be tempted to think that motifs emerge as self-organized modules that encode domain-specific information, is not sharply clear the measure in which different motifs stem from the diversity in learning environments and initial conditions or such diversity is a noisy by-product of the learning dynamics itself. 

\section{Discussion}

In complex networks, individual units by themselves do not accomplish any particularly relevant function, because it is the coordinated arrangement of groups of units (i.e. their \emph{interactions}) that allows for the emergence of system-level, macroscopic properties \citep{wuchty2003}. In the present work, we thus explored how information processing in deep networks might emerge as a combination of simple network motifs.

The key observations that 

\begin{itemize}
	\item Larger motifs may be seen as arrangements of smaller motifs, for example ``Diamonds combine to form multi-layer perceptron motifs'' \citep{alon2007},
	\item These smaller motifs arrangement gives rise to more complex computation: ``Adding additional layers can produce even more detailed functions in which the output activation region is formed by the intersection of many different regions defined by the different weights of the perceptron'' \citep[page~113]{alon2006} and
	\item Domain representation is carried out by the composition of subsequent non-linear modules, which ``transform the representation of one level (starting with the raw input) into a representation at a higher, slightly more abstract level'' \citep{lecun2015}
\end{itemize}

may indeed suggest that in deep neural networks the learning dynamics may rely on the same mechanism that in transduction networks. In particular the layer-wise character of the information synthesis process is appreciable both in transduction networks and neural networks. 

One might thus hypothesize that network motifs form spontaneously for the purpose of information processing and synthesising, so that each module deals with a small number of input features and subsequently motifs that lay deeper in the system deal with few activity signals of previous neurons. High-level features might thus be abstracted in a layer-wise and motifs-wise fashion. However, this view is not straight-forwardly supported by our analyses over six-nodes motifs, where the most common motifs did not have a clear relationship with those detected by the four-nodes analysis.

As a last remark, it is worth recalling the role of initialization schemes. As explained above, some strategies impress the initial significance landscape a sharper imprinting. In this initial rougher profile the motifs \emph{diamond} and \emph{bi-parallel} turn out to be the most significant. Thus the fact that in the case of orthogonal matrices and Xavier schemes the learning speed is larger may be put in relationship with the presence of these structures. The normal initialization scheme renders a flatter initial landscape, due to the smaller values sampled. Then the only relevant topological structures surfaced in the initial configuration are those suspected of being favored by the structural bias. The environment may be considered to be learned once relevant information processing structures come to develop. Clearly, if a scheme provides the initial configuration with a preventive signature of such structures, learning will be much faster.

\section{Final remarks and further improvements}

Besides the interesting results that we presented, there are some aspects that could be improved. As a first instance, feed-forward neural networks are the workhorse of deep learning, but there also exist models with bidirectional connectivity \citep{salakhutdinov2009deep}, which might allow for the emergence of a much richer variety of motifs.

Secondly, it would be useful to better investigate whether particular motifs might emerge simply as a consequence of some combinatorial bias induced by the design topology of the multilayer-perceptron itself. If so, some of the significant four-nodes motifs we detected might appear because of an unavoidable initial imprinting due to the topology of the model, and not necessarily because of their critical role in information processing. In addition, also the case of six-nodes groups would deserve some further inspection, in order to better characterize the functional role of these more complex network motifs.

Other than that, it could be interesting to enforce an initial landscape in which the motifs observable are those that appear to be most rare, according to the analyses presented. In this way one could observe whether these structures are rejected by the system evolution, favoring those above discussed. It would be a clear sign hinting that the motifs that we have found are indeed important in the learning process in this kind of models, also refuting the hypothesis of the spoil stemming from the combinatorial bias, due to the connectivity of the nodes within the architecture of the network itself.  

Finally, referring to the Appendix~\ref{app:preprocess}, it should be noted that our simulations involved a certain number of parameters (e.g., the threshold used to binarize the weights) that were often set according to heuristics. The choice of mining a weighted or unweighted network also plays a role. Here we presented results related to a weighted analysis, which introduces variability in the discovered motifs. More specifically, motifs were composed of connections that fall in four categories: close to zero (i.e., not present), strongly positive, strongly negative and mildly positive/negative. In addition, one could either account for \emph{most significant} or \emph{most typical} motifs, the former being those instances in the same group of isomorphic graphs that display the largest significance and the latter are those which have a significance score that is closer to the average, over the same isomorphic group. For a broader account on the subject see \citep{piperno2018} and \citep{mckay2014}. 

\bibliography{bibfile}
\clearpage

\appendix
\section{Data sets generation}
\label{app:data}
\unskip

Some previous work are followed for the zero stage (\citep{kemp2008} and \citep{saxe2019}). A difference is that in these cited publications, synthetic data consists of categories, and the learning system should guess each item's feature. This leads to a difference in the covariance structure (cfr. Figures~\ref{fig:tree_cov} and \ref{fig:clus_cov} above and, for example, Figure~9 in \citep{saxe2019}), and is due to the fact that, for example in the binary tree data structure, in the present case correlation patterns tie together, in some extent, all of the nodes in the binary tree.

Each node is associated with a feature (recall, a random variable $x_j$ that is one entry of the random data vector $\bm{x}$), whilst class labels are assigned according to whether a data item matches some of the previously created data vectors in the case of the binary tree. In the case of independent clusters, the category is assigned according to which one of the independent clusters is selected, see below.

In the real world, data often come as rows of a so-called \emph{design matrix}. Each one datum is then an array of some \emph{features} characterizing the observation. Each one of these features is a random variable, distributed according to some unknown distribution. In this spirit, the data set can be characterised by a multivariate probability distribution. 

An interesting way to represent multivariate distributions is provided by probabilistic graphical models (PGMs). These models represent the causality of the random variables involved by means of a graph: Nodes encode random variables, while edges encode the relationships that tie these variables together (Chapter 16 of \citep{goodfellow2016}). In this language, the sets of synthetic data that may be interesting for the sake of the present work can be formalised as PGMs and also by this representation the statistical structure may be more evident.

\subsection{Binary tree data set}

\begin{figure}[htb]
	\centering
	\includegraphics[width=0.65\textwidth]{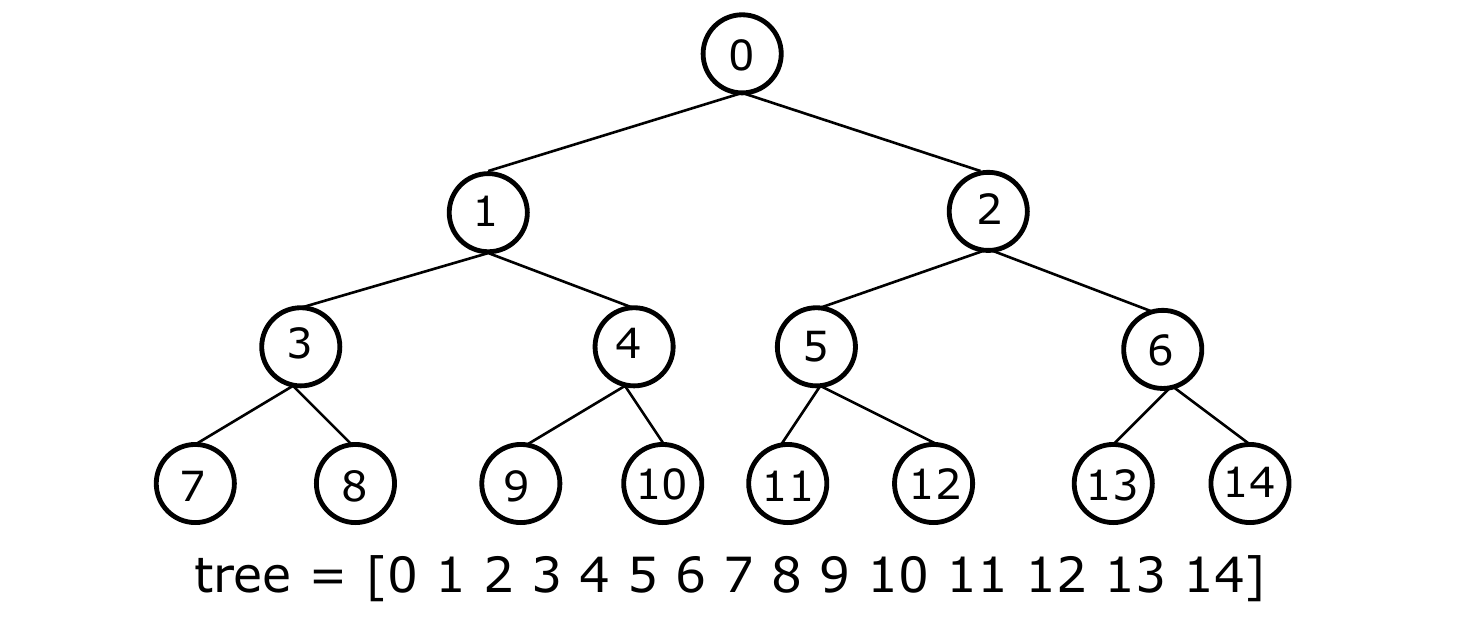}
	\caption[Binary tree data generating structure.]{Binary tree data generating structure. Note that the tree data structure is efficiently and easily represented computationally as a linear array. The left and right children of a given node $i$ are $2i + 1$ and $2i + 2$ respectively, with $i = 0, \dots, N-1$.} 
	\label{fig:tree}
\end{figure}

The first example is the binary tree data generating structure. The root node is a random variable, which attains one among the values $\{ -1,+1\}$ with equal probability $p = 0.5$. According to the outcome of such random variable, the children inherit the $\pm 1$ value according to some probabilistic decision rule and in the same fashion the children of the children, and so forth down the dynasty, see Algorithm~\ref{alg:singleFeature}. 

The user sets the depth $D$ of the tree to be created. A data sample is the collection of the $N = 2^{D} - 1$ random variables that constitute the tree structure. An advantage of the PGM representation is that it renders graphical visualisation ease: Data are often many-dimensional, i.e. points in a $N$-dimensional space.

In this case, the collection of $M$ of such vectors could be thought of as an ensemble of living species. The root node determines whether one item (pattern, data example) can move or not. The children of the root node determine whether \textit{if it moves, does it swim?} or \textit{if it does not move, does it have bark?}, and so forth. The levels deeper in the tree structure, bear more information about the data items. In the following, it is shown how the choice of a particular level resolves in the presence of more or fewer classes. As one considers the leaves level, then all of the nodes of a tree (i.e. all of the features in a pattern) must equal, for two items to belong to a given class. 

On the other hand, if one considers a shallow level in the tree structure, the nodes which must equal for two data vectors to belong to the same class, are all the nodes up to the last node of the level considered, plus those of the next level. For convenience, the root node lays at level $1$. To explain why it is in order to consider such nodes, refer to Figure~\ref{fig:tree}. Assume that one wants to gather in a class all the samples that \emph{move and swim}. Then the sample has, in first place, to actually \emph{move}, that is the root node must have the value $+1$, that means that its left child has value $+1$ as well and the right child and its dynasty inherit the $-1$ value. Then we should consider also the outcome of the stochastic inheritance of the $+1$ value from the node 1 to nodes 3 and 4. Based on this trial, we know whether the $+1$ value is attained by node 3 or 4. Node 3 encodes the answer to the question \emph{since the sample moves, does it swim?} by means of the value $+1$ which means \emph{yes}. Hence, to say whether two samples belong to the same \emph{moving and swimming animals} super-class, we should check the equality of the nodes \emph{almost} up to nodes 3 and 4. In practice, it is easier to check level-wise, thus two samples belong to the same class if all the nodes up to those of the next level match. Next level means next with respect to the level of detail one wants to inspect. In this example the level is $2$. All the subsequent nodes could in principle attain different values but this does not matter. If one wants to differentiate living things based on the fact that such items \textit{can move or not}, what matters is the value attained by the root node. Then whether two items are respectively a whale or a deer, this does not affect the belongingness to the \textit{living thing that can move} super-class. In contrast, if one has to differentiate \textit{living thing that moves} based on the fact that such an item \textit{does swim or not}, then a further level of detail is needed. Such a finer granularity is encoded by the values the nodes of the next levels attain. If the left children of the root node happen to inherit its $+1$ value, that means that other than \textit{being a moving living thing}, that item \textit{does swim}. Therefore, the second tree level encodes this subsequent level of detail. The more detail is embedded (the higher level is chosen), the more the possible classes the data examples may in principle belong to.

The rationale behind such a data generator is first and foremost related to its transparency and statistical structure clarity: There is no real-world consistency in such data, but in this fashion, it is easy to perform classification on them. As explained below, one single pattern generation happens to be a value diffusion down to the tree branches. In this way, one ends up with a $N$-dimensional binary array, in which many of the slots bear the $-1$ value. The $+1$ values, on the other hand, lays in correspondence of the slots associated with those nodes which happen to represent a positive answer to the distinction question associated with that node. Consistently with the discussed example: If the living thing encoded in such a $N = 15$ dimensional vector is a moving thing (roughly speaking, an animal), then the root node has the $+1$ value, which in turn means that the 0th slot in the data vector has such value. If this is a water animal, it swims, then the left child of the root node has inherited the $+1$ value, then the slot 1 in the data vector has the value $+1$ and it implies that the right child of root inherited the value $-1$, so the slot 2 of the data vector has the value $-1$. Assume further that other than swimming, this animal \textit{is not a mammal}. Then the left child of the 3-labelled node has inherited the $-1$ value and this same value is found in slot 7 of the data vector. It means that the $+1$ value is inherited by the right child of node 3, then in the final data vector the $+1$ value appears in slot 8. 

At the end of the day, the final data vector is made up by $-1$s, except for these said slots, where the $+1$ value ended up in, encoding the positive outcome of those criteria associated with the respective nodes. The terminal (leaves) level could be imagined as the \textit{one-hots} stratum, that is: all of the leaves attain the $-1$ value, except for one single leaf, where the $+1$ got to settle, as a consequence of the (stochastic) outcome of all the aforementioned decisions. This lonely $+1$ determines the final category in which the data vector fits in, as one sets the leaves level to be the distinction granularity. In such case, for two vectors to belong to the same class, it must be that all of the features equal. Otherwise, it could in principle be that a whale, echoing the previously discussed example, has the root node positive, but in another data row, it could be negative. This would mean that a whale is a \textit{not moving living being} that \textit{swims}. So, in the label generation stage, one shall differentiate according to all the nodes of the level under consideration and all of their ancestries.

\subsubsection{Single pattern generation}

One pattern is the collection of \textit{all} the node values of the array-represented tree (that entity formerly dubbed a \textit{data vector}).
As an example, to the non-leaves nodes are associated decision rules, intended to discriminate samples (e.g.: \textit{does the object move?}, which can be answered with \textit{yes} or \textit{no}, $\pm 1$, is the primal decision rule, i.e. axis along which one can set distinctions). The initial value of the root node is inherited and eventually flipped according to probabilistic decision rules with respect to a fixed probabilistic threshold $\varepsilon$.

In this spirit, referring again to Figure \ref{fig:tree}, the (non-leaves) nodes ranging from 0 to 6 encode decision rules, (leaves) nodes indexed with $i = 7,\dots\,14$ represent the final details about a sample which are not important for the sake of its classification. The following criteria are implemented:

\begin{enumerate}
	\item The probabilistic threshold is fixed a priori. The smaller its value, the less variability in the data set.
	\item Root attains the values $\pm 1$ with probability $p = 0.5$.
	\item Root's children attain values $+1$ or $-1$ in a mutually exclusive fashion. The following convention is adopted: \textit{if the root node attains the value $+1$, then the left child inherits the same value. Else, the left child attains the value $-1$ and the right child has assigned the value $+1$}. 
	\item From the third level (children of root's children), the progeny of any node that has value $-1$ also has to have $-1$ value. On the other hand, if one node has value $+1$, its value is inherited (again mutually exclusively) by its children according to a probabilistic decision rule. 
\end{enumerate}

The aforementioned probabilistic decision rule is a Metropolis-like criterion: Sample a random variable $p \sim U([0,1])$, then, given the probabilistic threshold $\varepsilon$,

\begin{itemize}
	\item If $p > \varepsilon$, the left child inherits the $+1$ value, and the right child, alongside with its progeny, assume the opposite value;
	\item Else, is it the right child to assume the value $+1$.
\end{itemize}

\subsubsection{Complete data set}

Repeating the above procedure $M$ times, one ends up with a data matrix $\bm{X} \in \{-1, +1\}^{M \times N}$, i.e. each row of $\bm{X}$, $\bm{x}^{\mu}$, $\mu = 1 , \dots , M$, is one single $N$-dimensional data vector, in the same terminology as above, that is a $N$-featured data vector (one pattern).

To complete the creation of a synthetic set of data, one needs the \textit{label} associated with each one of the data items. Here the choice of the probabilistic threshold $\epsilon$ turns out to be crucial. The higher this quantity, the more the total number of different classes the data example may fall into. On the other hand, if $\epsilon$ is small enough, there is a low probability of flipping a feature value, then it is more likely to observe repeatedly the same configuration. 

To create the labels, encoded as \textit{one-hot} activation vectors, one arbitrarily assumes the identity matrix to be the labels matrix. Then the whole data set is explored in a row-wise fashion. Since the data set has a hierarchical structure, it is possible to select the granularity of the distinction made in order to differentiate patterns in different classes. It depends on the choice of a level in the binary tree: If the level chosen is high (far away from the root node) then one ends up with a fine-grained distinction. On the other hand, if the chosen level is low, the distinction is made according to \textit{super-classes}, e.g. whether a given object \textit{can move}. The finer the granularity, the more detailed the distinction between patterns. Obviously, in this latter case, the data set exhibits a greater number of distinct classes. See the discussion above.

By this observation, the label matrix is created according to the level of distinction chosen. The node values to be considered (i.e. the entries of each $\bm{x}$ data vector) are all those that encode the values of the nodes up to the last one of the level selected. Referring again to the tree in Figure \ref{fig:tree}, if it suffices to identify the \textit{move or not} distinction, then one could safely check both tthe root node only or the root node with its children, that is nodes $\{0,1,2\}$, because the inheritance from the root node to its children is conventionally based on the root value solely. But if one wants to consider whether an object \emph{can move} alongside with the further \textit{if it moves, does it swim?} and \textit{if it does not move, does it have bark?} distinctions, then one should consider also the children nodes of node 1, that is the answer to the decision rule asked by node 1. Hence to determine whether two data items fall in that same category, we check that all the first $2^{L+1}-1$ nodes have the same value. Here $L = 2$, in fact we consider nodes $i \in [0, 2^{L+1}-1] \equiv [0, 7] = \{0,1,2, \dots, 6\}$.

By thus doing the data set is generated. The matrices $\bm{X}$ and $\bm{Y}$ are saved to a proper data structure which can be easily managed by the program that implements the artificial neural network described in the main text.

\begin{algorithm}
	\begin{algorithmic}[1]
		\caption{Binary tree. Single feature generation}
		
		\State Compute $N = N_{\text{leaves}}$, $n = N_{\text{not leaves}}$. $M$ is a free parameter \\
		\State \texttt{tree} = $\bm{0}^{N}$ \\
		\State Define a small $\varepsilon \sim O(10^{-1})$ as probabilistic threshold \\
		\State Value of root $\eta^{(0)} \sim U(\{-1, \, +1\})$ \\
		\If {Root node has value $+1$} \\
		\State The left child inherits the value $+1$ \\
		\State And the right child inherits the value $-1$ \\
		\Else \\
		
		\State The left child inherits the value $-1$ \\
		\State And the right child inherits the value $+1$ \\
		\EndIf \\
		
		\For {All the other nodes indexed $i = 1,\dots,n$} \\
		
		\If {Node $i$ has $+1$ value} \\
		\State Sample $p \sim U([0,1])$ \\
		\If {$p > \varepsilon$} \\
		\State Left child of $i$ = $+1$; Right child of $i$ = $-1$ \\
		\Else \\
		\State Left child of $i$ = $-1$; Right child of $i$ = $+1$ \\
		\EndIf \\
		\Else \\
		\State Both the children of $i$ inherit its $-1$ value \\
		\EndIf \\
		\EndFor \\
		
		\State $\bm{x}^{\mu} \leftarrow $ values generated, $\quad$ $\mu = 1 , \dots, M$ \\
		
		\label{alg:singleFeature}
	\end{algorithmic}
\end{algorithm}

\begin{algorithm}
	\begin{algorithmic}[1]
		\caption{Binary tree. \textit{One-hot} activation vectors, i.e. labels}
		
		\State Choose level of distinction $L$ \\
		\State $\bm{Y} = \mathbb{I}$ \\
		
		\For {$\mu = 1, \dots, M$} \\
		\For {$\nu = i, \dots, M$} \\
		
		\If {the first $2^{L+1}-2$ entries of $\bm{x}^{\mu} \text{ and } \bm{x}^{\nu}$ equal} \\
		
		\State $\bm{y}^{\nu} \leftarrow \bm{y}^{\mu}$ \\
		\EndIf \\
		\EndFor \\
		\EndFor \\
		
		\For {$i = 1, \dots, N$} \\
		\If {$\bm{Y}[:,i] \text{ equals } \bm{0}^{N}$} \\
		\State Eliminate column $i$ of $\bm{Y}$ \\
		\EndIf \\
		\EndFor \\
		
		\label{alg:labelMatrix}
	\end{algorithmic}
\end{algorithm}

\subsection{Independent clusters data set}

The generation of the second data set is performed as follows: Generating some cloud of points distributed according to a bivariate Gaussian distribution, with means spread apart and covariances sufficiently small, in such a way that the points of different groups do not overlap with the others. The 2-dimensionality has, of course, nothing to do with the number of features, which as said before is the total number of points generated, that is the nodes of the probabilistic graph representation. This 2-dimensionality serves solely to draw the PGM and subsequently to partition the graph. 

Once points are generated, are turned in a fully connected graph, i.e. create edges between each pair of nodes. In the spirit of the \textit{simulated annealing} algorithm, here it is imagined that such a fully connected graph is a sort of mineral structure, and it is in order to increase the temperature, to simulate a melting process that destroys some of the over-abundant edges, according to some metric, for example, the distance between points. For this reason, it comes handy the 2-dimensional representation: Distance is simply the norm of the vector from a node to another. The distance for which the edge is removed is temperature-dependent: the higher the temperature, the shorter the maximum edge length allowed. At the end of this \textit{simulated melting} process, it is expected the graph to exhibit some independent components, provided the melting schedule is properly set. Moreover, these independent groups are not fully connected within themselves. The melting schedule is designed in a way to remove some of these intra-edges. This simulates the random variables of each group not to be dependent on all of the others in the same could. Note that, unlike how exposed in \citep{kirkpatrick1983}, in this melting simulation there is not, strictly speaking, a \textit{optimization} perspective inasmuch what matters is the removal of some edges. The physics of the procedure could be improved.

\begin{figure}
\centering
\includegraphics[width=0.7\textwidth]{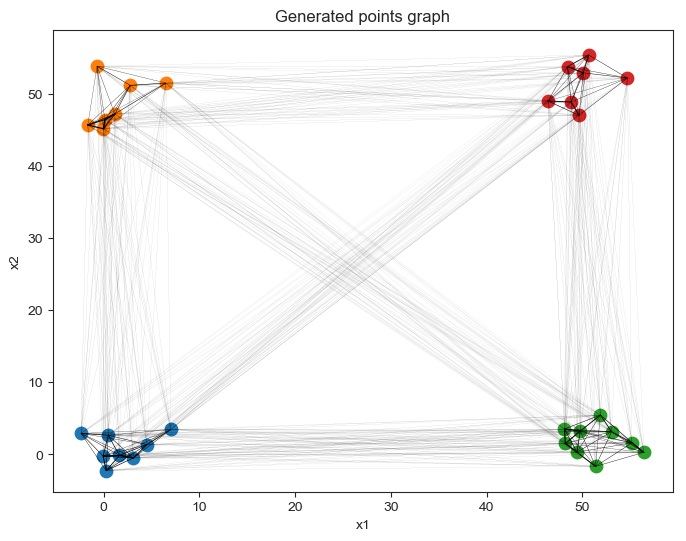}
\caption{First stage of the data set generation for the independent clusters structure.}
\end{figure}

\begin{figure}
\centering
\includegraphics[width=0.7\textwidth]{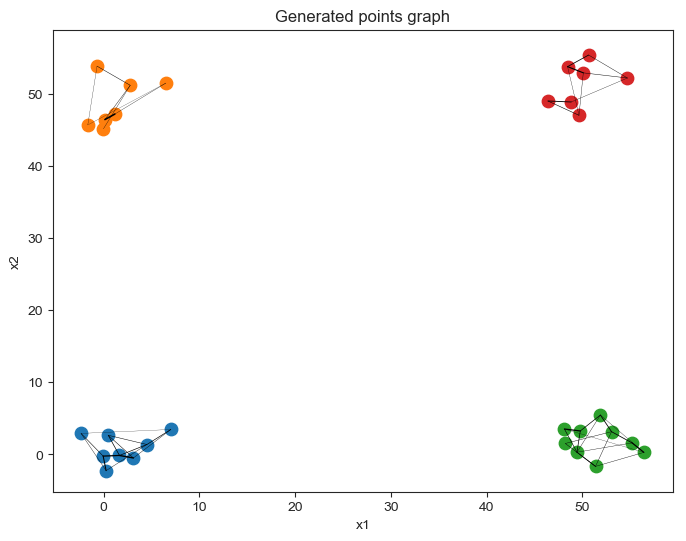}%
\caption{Second stage of the data set generation for the independent clusters structure.}
\label{fig:clus_graph_plots}
\end{figure}

\begin{algorithm}
	\begin{algorithmic}[1]
		\caption{Independent clusters. Simulated melting to \textit{partition the graph}}
		
		\State Choose the number of classes $N_C$ \\
		\State Set $\bm{\mu}^{(k)} \in \mathbb{R}^2$, $\bm{\Sigma}^{(k)} \in \mathbb{R}^{2\times2}$, $k = 1, \dots, N_C$ \\
		\State Generate $\bm{X}$ s.t. $\bm{x}_i \sim \mathcal{N}(\bm{\mu}^{(k)}, \bm{\Sigma}^{(k)})$, $i = 1, \dots, M$ \\
		\State Include the indexes of the points generate in a list, which is the set of the vertices $\mathcal{V}$ of the graph $\mathcal{G}$ \\
		\State Fully connect the vertices to form a fully connected graph and group the vertices and the set of the edges $\mathcal{E}$ in the graph data structure, $\mathcal{G} = \{\mathcal{V}, \mathcal{E}\}$. \\
		\State \textbf{Note} that since 2-dimensional coordinates will be useful, $\mathcal{V}$ is a dictionary of keys (nodes indexes $i = 1, \dots, M$) and values (list with the point coordinates, $(x_i^{(1)}, x_i^{(2)})$). \\
		
		\For {$T$ increasing} \\
		
		\For {All the edges $e = 1,\dots, |\mathcal{E}|$} \\
		
		\If {Length of edge $e > \frac{1}{T}$ (for example) } \\
		
		\State Remove edge $e$ \\
		\EndIf \\
		\EndFor \\
		\EndFor \\
		
		\State Plot the remaining edges and check if only independent fully connected components have survived. \\
		\label{alg:simulatedmelting}
	\end{algorithmic}
\end{algorithm}

\begin{algorithm}
	\begin{algorithmic}[1]
		\caption{Independent clusters. Single pattern generation}
		\State Here $i$ indexes a single random variable. This kernel is used as many times as the number of samples the user wants to generate. $\bm{x}$ is the whole data item, initialised with each slot set to $-1$. \textbf{Note}: in the data set actually generated, the value of the nodes are set to their topological orders, with no ancestral sampling implemented. \\
		\State Set $\bm{x}$ = $\{-1\}^N$ \\
		\State Sample $L \sim \mathcal{U}(\{ 1, \dots, N_c\})$ \\
		\For {all the vertices $i = 1,\dots, n_k$ in cluster $L$} \\
		\If {Topological Order of $i$ is 1} \\
		\State $x_i \sim p(x_i) \sim \mathcal{N}(0, k_i^{-2})$ \\
		\Else \\
		\State $x_i \sim p(x_i) \, \prod_{j \in \text{Ancestors}(x_i)} (4\pi \, k_j^2)^{-1/2}\, \exp \left(-\dfrac{1}{2} \dfrac{x_j^2}{k_j^2} \right)$ \\
		\EndIf \\
		\EndFor \\
		\State $\bm{y}_i = \text{one-hot}(L)$  \\
		
		\label{alg:clus_spg}
	\end{algorithmic}
\end{algorithm}

\subsubsection{Single pattern generation}

Once the independent clusters come to form, it is to assign each of the nodes a topological ordering in such a way to perform the ancestral sampling \citep[Chapter~16]{goodfellow2016}. Since the graphs are directed, in the edges data structure created each edge is in the form of a couple $(i,j)$, i.e. edge from node $i$ to node $j$. Then if one node appears only on the left slot of such representation, it has topological order 1, in that no edge ends up at that node. Conversely, each node appearing on the right has almost one ancestor. For each edge then, each right node is saved to a proper data structure, and it is kept track of the ancestors of each node. In this way, it is possible to assign both the topological order and to keep a list of all the ancestors. It will be useful in the stage of sampling to dispose of such a list. 

As a zero model however it is done as follows: A data item is initially initialized with all the features values of $-1$. Since each vertex in the graph encodes a feature, and the belongingness of each vertex to a group is an information known from the points generation stage, an integer ranging from 1 to the number of classes $N_c = 4$ is sampled uniformly. The nodes corresponding to this label number are assigned different values, according to their topological order. This is trivial to do since for each vertex belonging to the selected group one simply puts in the corresponding slots in the data vector the topological order of such vertices. 

A further improvement could be rather this approach: once a label is sampled, one could sample from the distribution $p(x_i)$, for the vertices with topological order 1 in that cluster. The values associated with nodes having topological order 2 is still sampled from that distribution, but must be conditioned to the values sampled for their ancestors (nodes of order 1), i.e. $p(x_i \, | \, \mathrm{Ancestors}(x_i))$. This is explained by recalling the very purpose of graphical models: to show (even graphically) the \textit{causality} of the random variables involved. As distribution, it could be chosen a Gaussian with mean zero and variance proportional to the degree of that node. Gaussian is believed to fit since nearby features are expected to have similar values \citep{kemp2008}, but differently from this work, here one does not generate the features vector, hence sampling from the multivariate Gaussian having zero mean and variance dependent on the inverse of the Laplacian matrix of the graph. Here it suffices to sample a value for a single node, and hence the degree of a node could be a good compromise, being such quantity one of the ingredients of the Laplacian).

To sample from the conditional $p(x_i\, | \, \mathrm{Ancestors}(x_i))$ the following rationale may be implemented: The distribution is referred to all the nodes up to $i$, then could be viewed as a multivariate distribution. Then a value is sampled from that multivariate distribution, but keeping constants the values of the random variables sampled yet. As an example: Assume that node 3 of cluster 1 is to be assigned the value $x_3$ and that $\mathrm{Ancestors}(x_3) = [1,2]$. Then the distribution to sample from is 

\begin{equation}
\displaystyle p(x_3 \, | \, x_1, x_2) \sim \exp \left( - \frac{1}{2} (x_1, x_2, x_3)^T \, \bm{\Sigma}^{-1} \, (x_1, x_2, x_3) \right)
\end{equation}

with $\bm{\Sigma} = \mathrm{diag}(k_i)$, $i = 1, \dots, \ 3$, being $k_i$ the degree of node $i$. The above formula may be broken in products, owing to the fact that the variance matrix is diagonal, that is 

\begin{equation}
\begin{array}{r c l}
\displaystyle p(x_3 \, | \, x_1, x_2) &\sim& \exp \left(-\dfrac{1}{2} \dfrac{x_1^2}{k_1^2}\right) \, \exp \left(-\dfrac{1}{2} \dfrac{x_2^2}{k_2^2}\right) \, X_3 \\[3mm] 
X_3 &\sim& \mathcal{N}\left( 0, k_3^{-2} \right)
\end{array}
\end{equation}

the first two factors being the values that the Gaussian probability density function attains at the values sampled for the ancestors $x_1$ and $x_2$ and the third factor is the value sampled from the Gaussian having zero mean and variance $k_3^2$.

\subsubsection{Complete data set}

This procedure is repeated many times as specified by the user. Here a good number is, as in the case of binary tree, $M = 2000$ items. In the complete data set hence one has features in which the only values not being $-1$ lay in correspondence of the indexes of the data array that match with the nodes of the graph that belongs to the category given by the label of that feature. Labels are again \textit{one-hot} vectors. For example, assume that the first cluster is selected. If this first cluster comprises the vertices ranging from 1 to 5, where node 1 has order 1, 2 and 3 have order 2, 4 has order 3 and five has order 4, then that data item has values $[1, 2, 2, 3, 4, -1, \dots, \ -1]$ and the corresponding label is $[1, 0, \dots\, , 0]$.

\unskip
\section{Pre-process}
\label{app:preprocess}

\begin{figure}
\centering
\includegraphics[width=\textwidth]{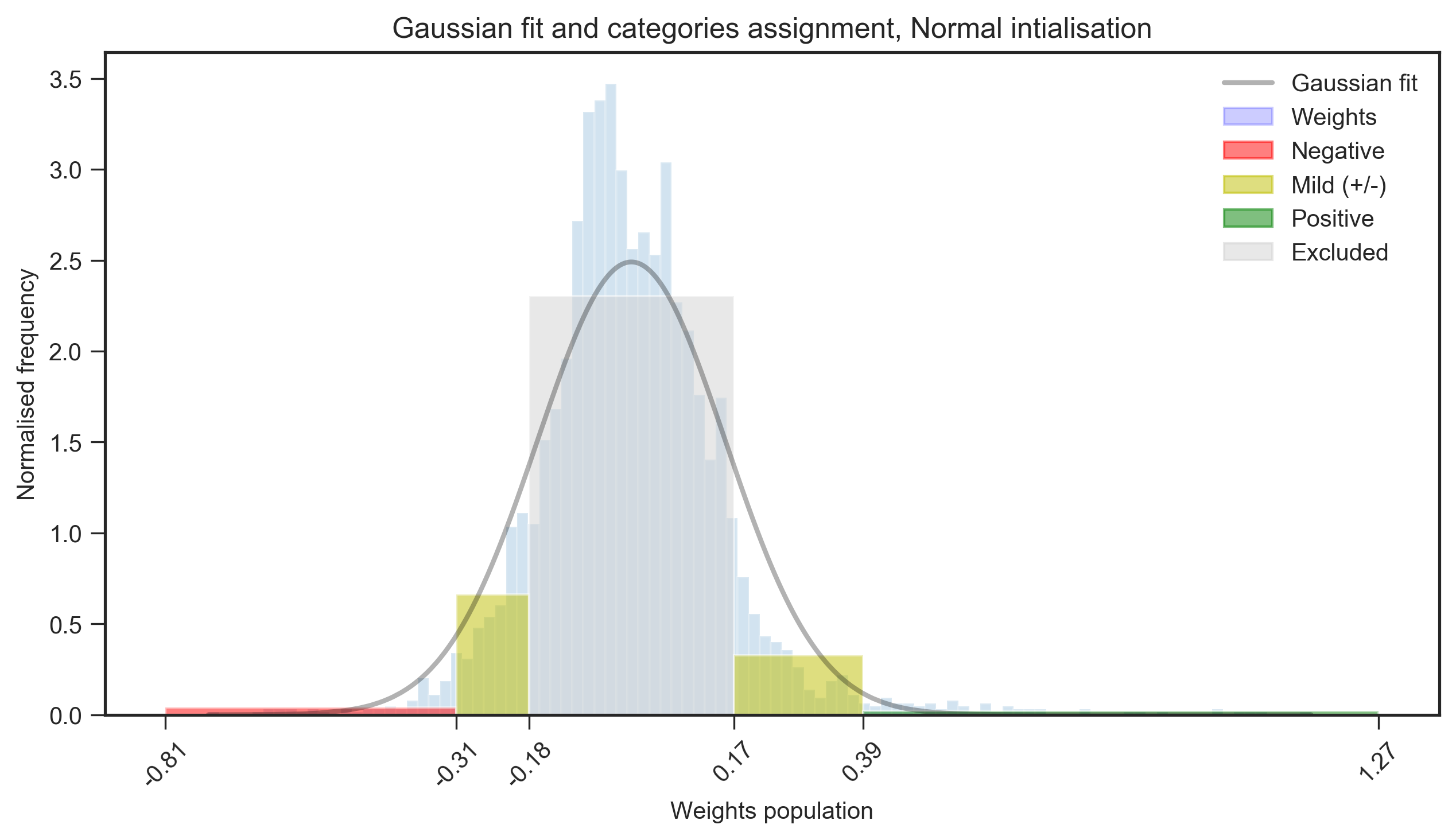}
\caption{Gaussian fit of the weights population and the respective subdivision histograms for the entire weights matrix.}
\label{fig:gauss_fit}
\end{figure}
	
\begin{figure}
\centering
\includegraphics[width=\textwidth]{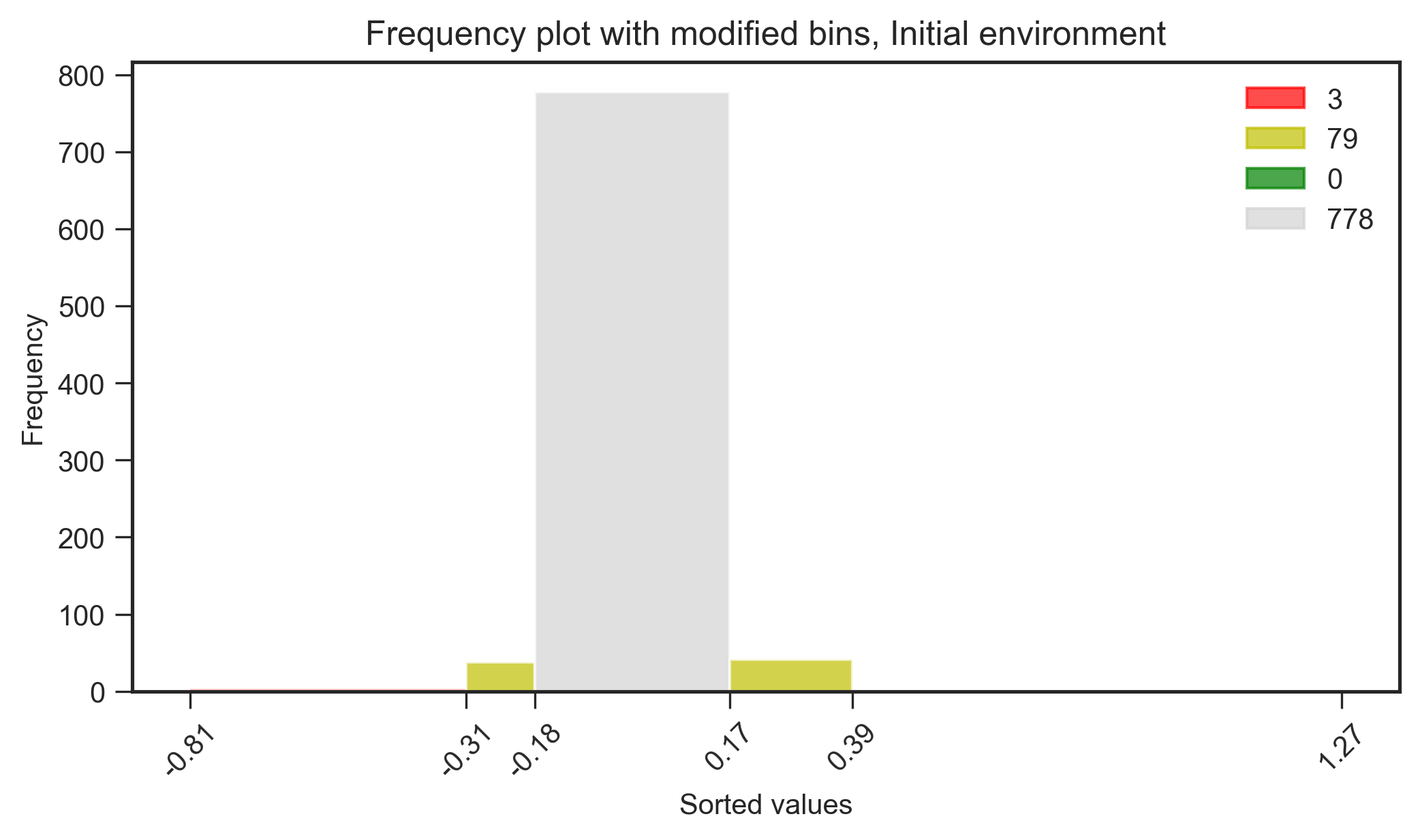}
\caption{Gaussian fit of the weights population and the  subdivision histograms for the initial configuration.}
\label{fig:init_hist}
\end{figure}
	
\begin{figure}
\centering
\includegraphics[width=\textwidth]{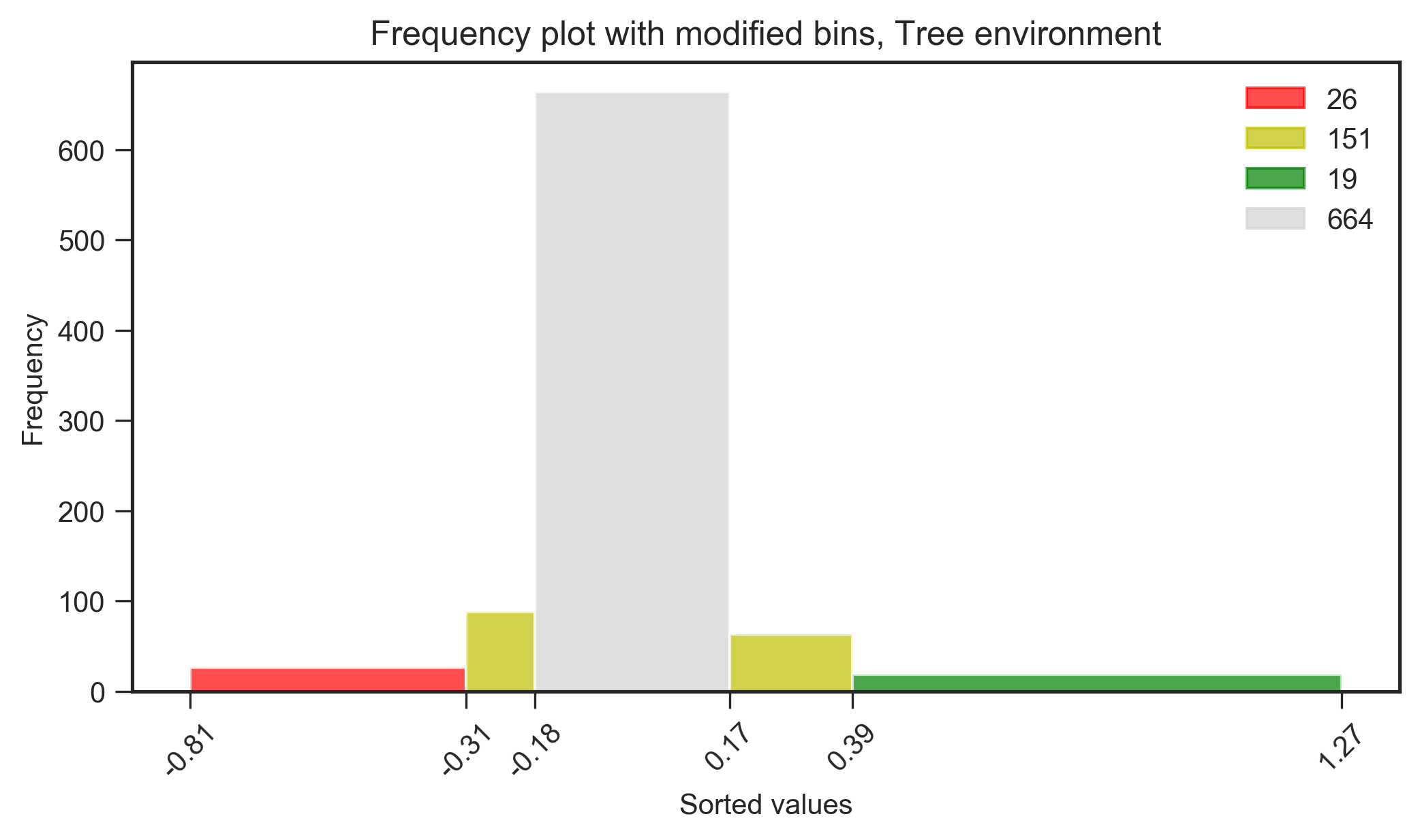}
\caption{Gaussian fit of the weights population and the  subdivision histograms for the tree data set.}
\label{fig:tree_hist}
\end{figure}
	
\begin{figure}
\centering
\includegraphics[width=\textwidth]{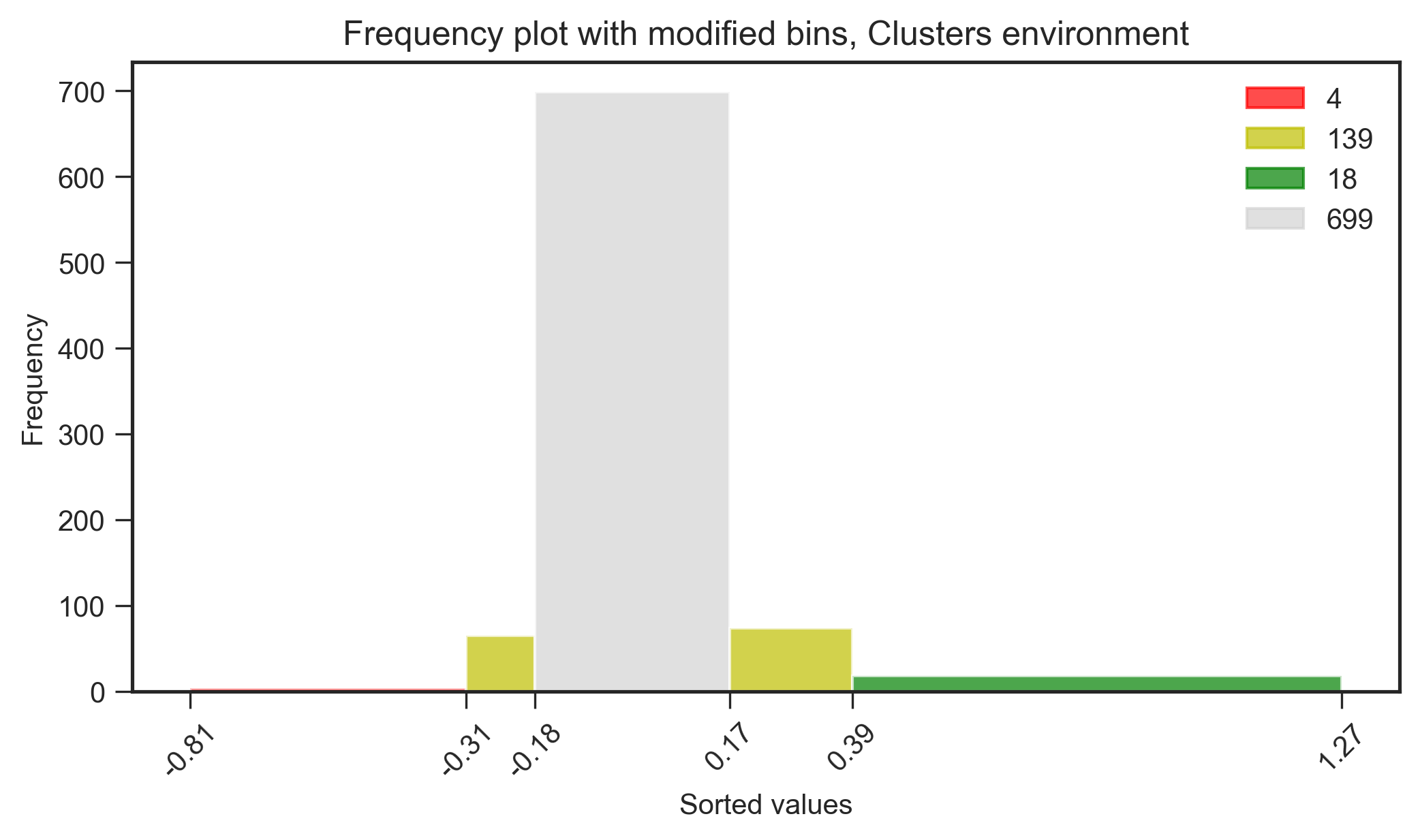}
\caption{Gaussian fit of the weights population and the  subdivision histograms for the clusters data set.}
\label{fig:clus_hist}
\end{figure}

Once the model is trained, it is necessary to pre-process the parameters data structures to subsequently use the motif detection tools. The FANMOD (FAst Network MOtif Detection) software has been utilised. In the case of unweighted networks, one simply sets all the connection strengths to $1$ (those that fall in the non-zero category, see below), contrarily, weights must be discretised, in that motif mining software deal with colored networks, which means that edges tags must be categorical. The categories supported by the software are limited (maximum seven). 

The following heuristic is followed: One wants the categories characterising the edges to be: positive, negative, neutral, or mildly negative/positive. For each initialisation scheme, the model parameters are gathered in one single array, so to consider for the maximum range of such parameters. Accounting for all the weights with no threshold would resolve in a redundant number of motifs instances detected in the mining stage. So, to preventively prune such over-necessary information while keeping some weights that are \textit{lesser} than some others, in an absolute sense, the closest to zero connection weights are removed. Categories are assigned based on a decomposition of the tail of the distribution of the model parameters. As shown in Figures~\ref{fig:gauss_fit}, \ref{fig:init_hist}, \ref{fig:tree_hist} and \ref{fig:clus_hist}: Perform a Gaussian fit of the entire weights population, collected for all the model configurations (initial, trained with both the data sets). While some other viable alternatives to the normal distribution may fit even better, it is thought to be a sound zero case. The exclusion interval is the set of those weights $\tilde{w}$ for which the corresponding value of the fitted probability density function $p(\tilde{w})$ is greater or equal that $0.55 \times \max_w p(w)$. This choice is a \emph{free parameter}. This precise threshold value turned out to be a good trade-off between inclusiveness of small weights and computational time in the execution of the FANMOD program. The survived pieces of the probability density function support are divided to render the weights categories as follows: $1/5$ of the tails, namely that closer to zero, is assigned to the category of the \textit{mildly positive/negative} weights, while the outermost fringes of the tails are assigned to the categories \textit{positive} and \textit{negative}, according to the side of the probability density function. The 1/5 value is another free parameter. Figure~\ref{fig:gauss_fit} refers to the normal initialization scheme. Bottom panels display the weights spectra for the different model configurations discretized in categories according to the scheme outlined above. The numbers listed in the legends are the counts of how many weights values fall in the respective category. Note that in the initial configuration the extreme fringes enjoy a lesser presence of connections.

Histograms depicted give a graphical explanation of this heuristic: The central bin, the most populated, is not taken into account in the subsequent analysis. The extreme tails are the positive and negative edges strengths values. The two intermediate bins, between the central one and the extreme ones, are those set to be the neutral weights.

In this way, the motifs featuring values falling in the central bin can be preventively discarded in further analyses in that they are not thought to be relevant. One has at this point the graph representations amenable to the motifs mining program. Whether colors shall be accounted for or not, it absorbs the input file formatted according to the above-discussed partition and performs the analyses. If the analysis to be carried out is on an unweighted network, then all the connection strengths are set equal to $1$, except those related to the central bin items, which are not accounted (no connection).

\end{document}